\documentclass[12pt,preprint]{aastex}

\shorttitle{RADIATION SPECTRA AND POLARIZATION IN MAGNETAR BURSTS}
\shortauthors{Niemiec, Bulik}
\usepackage{natbib}

\begin{document}

\title{RADIATION SPECTRA AND POLARIZATION IN MAGNETAR BURSTS}

\author{Jacek Niemiec}
\affil{Department of Physics and Astronomy, Iowa State University, Ames,
IA 50011, USA}
\affil{and Institute of Nuclear Physics PAN,
 31-342 Cracow, Poland}
\email{niemiec@iastate.edu}
\and
\author{Tomasz Bulik}
\affil{Nicolaus Copernicus Astronomical Center, 00-716 Warsaw, Poland}
\email{bulik@camk.edu.pl}

\begin{abstract}
We present Monte Carlo simulations of radiative transfer in magnetar
atmospheres. We include the effects of vacuum
polarization, electron and proton scattering, and free-free absorption.
Simulations are performed for the atmosphere model with the
magnetic field perpendicular and also tilted with respect to the
neutron star surface, and  we show that the average spectrum
does not strongly depend on the orientation of the magnetic field.
We investigate the region of the parameter space where the vacuum
absorption-like feature appears  in the
spectrum and we analyze
the shape of the proton cyclotron line.
 Our results indicate that the existence of the vacuum polarization feature 
should be a general attribute of soft gamma-ray repeaters burst spectra, 
provided that the energy release takes place at the sufficiently dense 
region, and the atmosphere scaleheight is large  enough. We discuss the 
existence of  such a feature in recent observational data on these sources.

\end{abstract}


\section{INTRODUCTION}

Soft gamma-ray repeaters (SGRs) and anomalous X-ray pulsars (AXPs)
belong to a group of objects  called magnetars, since the main source of 
energy powering their radiation comes from the superstrong
magnetic field. 
The first observational suggestion of the superstrong 
magnetic field came from analysis of  the
March 5th burst properties  \citep{1979Natur.282..587M,1992AcA....42..145P}, 
and the basic model of magnetar emission 
was proposed by \cite{1995MNRAS.275..255T}.
The existence of superstrong 
magnetic fields has been inferred from the spindown 
measurements \citep{1998Natur.393..235K}.
The  AXP population exhibits a lot of 
similarities to the SGRs. Their spins and spindown rates 
are similar, and recently AXPs have also shown 
bursting behavior \citep{2002Natur.419..142G}.
For a review of the properties of magnetars see \citet{2000AIPC..526..763H}
and \citet{2001AIPC..599..219M}.

In the conditions prevailing in a magnetar atmosphere, the basic
contributions to the opacity come from the electron scattering,
free-free absorption, and the proton scattering.
The atmospheres should be   completely ionized during the bursts,
if they are composed mainly of hydrogen,
yet in the quiescence some non-ionized material can be present.
The density changes very rapidly since the gravitational 
acceleration is in excess of $10^{14}$cm\,s$^{-1}$. 
For typical magnetar fields,
the proton cyclotron frequency  $\hbar\omega_p=6.4 B_{15}$\,keV,
where $B_{15} = B/10^{15}$\,G, comes right in the
X-ray range.
Important additional effects arise from the polarization of the vacuum.
These effects for strong magnetic fields have been discussed
by several authors 
(e.g., Tsai \& Erber 1975; Ventura et al. 1979; Pavlov et al. 1980, for a 
review see, e.g, Pavlov \& Gnedin 1984; Meszaros 1992).

Modeling the atmospheres of strongly magnetized sources
is usually done  using the two normal modes description.
There are basically two methods to solve the radiative 
transfer equation: the finite difference scheme and the Monte Carlo
method. The finite difference scheme was
used to model the radiation spectra of X-ray pulsars
\citep[e.g.,][]{1985ApJ...298..147M,1992ApJ...395..564B}, 
and also  applied to the case of the superstrong
magnetic fields  
\citep[e.g.,][]{1996AIPC..384..907B,2001MNRAS.327.1081H,%
2001ApJ...563..276O,zan01}.
The Monte Carlo method was also used in modeling of 
accreting X-ray pulsar 
spectra \citep{1989ApJ...338..343W,1998ApJ...505..688I} and to the 
atmospheres with superstrong magnetic fields \citep{bul97}. The difference 
scheme method allows to  construct self-consistent models of static 
atmospheres, yet requires a careful choice of the grids in order to resolve 
the sharp resonances \citep[see, e.g.,][]{2003ApJ...583..402O}. Imposing 
the radiative equilibrium is easy in this approach.
Inclusion of the Comptonization effects 
is, however, much more difficult.
In the case when the 
magnetic field is tilted, not perpendicular to the surface,
an accurate solution of radiative transfer becomes extremely 
tedious within the difference scheme. The azimuthal 
angle is an additional degree of freedom, and the dimensions 
 of the matrices increase significantly. However, 
 approximate solutions 
can be obtained in the diffusion approach 
\citep[e.g.,][]{1995lns..conf...71P}. The Monte Carlo 
method allows to treat the Comptonization effects easily and 
permits to resolve sharp resonances. 
Imposing the radiative equilibrium constraints 
is much more difficult.
Past attempts to use the  Monte Carlo technique 
suffered from the limited statistics.
The method  also allows to investigate the tilted field case, as
the length of calculations does not increase strongly 
when the rotational symmetry is broken.
It has to be emphasized that 
a comparison of the two methods provides a cross-check 
to independently confirm the results.

The effects of the vacuum polarization in application
to the superstrong fields  have been first  
analyzed by \citet{bul97}, who 
solved the  radiative transfer using a Monte Carlo method. 
They modeled the atmospheres with the temperatures above $10\,$keV,
typical of the SGR bursts.
They have shown that  vacuum polarization 
leads to appearance of a broad  absorption-like  feature
in the region around  $\sim 5-15$\,keV 
in the presence of superstrong magnetic fields.
The vacuum feature in the spectrum appears because of density
dependence of the vacuum resonance energy. 
The radiative transfer is dominated by the 
extraordinary mode because of its  much lower 
cross-section in comparison to the 
ordinary mode.
The vacuum resonance then leads to creation of an opaque layer 
for the low-cross-section mode, which effectively 
changes the depth of  its photosphere. 
\citet{bul97} neglected the effects of proton cyclotron resonance.
In recent years,  the difference scheme solution of magnetized
atmospheres   has been applied to compute the spectra
of magnetars in a series of papers by 
\citet{2001ApJ...563..276O}, 
\citet{2001MNRAS.327.1081H,2003MNRAS.338..233H} and 
\citet{2003ApJ...599.1293H}.
They have mainly considered the case of emission 
in quiescence, with temperatures around $10^6$K. The proton cyclotron 
resonance was included and shown to appear as a rather prominent line. 
Proton cyclotron resonance has also been discussed by \citet{zan01} and 
\citet{2003ApJ...583..402O}.

In this paper  we present a Monte-Carlo-type calculations 
of the magnetar atmospheres. Our main goals are to 
 include and investigate  the effects of vacuum polarization 
the proton cyclotron resonance,  calculate the polarization of the 
outgoing radiation, and solve  the radiative transfer  in the case of an 
atmosphere with tilted magnetic field. In \S 2 we present the calculations of 
normal mode opacities and polarization vectors, and in \S 3 we describe the 
Monte Carlo procedure. \S 4 contains the presentation of the results,
which are discussed in  \S 5.

\section{RADIATIVE OPACITIES AND POLARIZATION}
Transfer of electromagnetic radiation in a magnetized medium can be 
conveniently described in terms of normal modes. The complex refractive 
indices of these modes $n_j$ are determined from the wave dispersion 
equation, which for the medium with the electric permittivity tensor 
$\epsilon$ and the magnetic permeability tensor $\mu$ becomes
\begin{equation}
\label{disp}
{\bf k \times [\mu^{-1} (k\times E)]} + 
\left(\frac{\omega}{c}\right)^2{\bf \epsilon E} =0,
\end{equation}
where $\bf E$ is the wave electric field. Solutions of equation (\ref{disp}) 
can be presented using notation introduced by Gnedin \& Pavlov (1974). In 
the coordinate system with the magnetic field along the $z$-axis, and the 
wavevector in the $yz$-plane at an angle $\theta$ to the magnetic field, we 
 define:
\begin{eqnarray}
\label{ni}
n_I & \equiv & 1 + \frac{1}{4}(\epsilon_{xx} + \epsilon_{yy}\cos^2\theta + 
\epsilon_{zz}\sin^2\theta - \epsilon_{xz}\sin 2\theta \\\nonumber
 & & - \mu^{-1}_{xx} - 
\mu^{-1}_{yy}\cos^2\theta - \mu^{-1}_{zz}\sin^2\theta + 
\mu^{-1}_{xz}\sin 2\theta),\\
\label{nl}%
n_L & \equiv & \frac{1}{4}(\epsilon_{xx} - \epsilon_{yy}\cos^2\theta - 
\epsilon_{zz}\sin^2\theta + \epsilon_{xz}\sin 2\theta \\\nonumber
 & & + \mu^{-1}_{xx} - 
\mu^{-1}_{yy}\cos^2\theta - \mu^{-1}_{zz}\sin^2\theta + 
\mu^{-1}_{xz}\sin 2\theta),\\
\label{nc}
n_C & \equiv & \frac{\rm i}{2}(\epsilon_{xy}\cos\theta + 
\epsilon_{xz}\sin\theta).
\end{eqnarray} 
Then, the complex refractive indices are 
\begin{equation}
\label{nj}
n_j = n_I \pm \sqrt{n_L^2 + n_C^2},
\end{equation}
where $j=1,2$ labels the normal modes. The (real) refraction coefficients 
of the normal modes are given by the real part of $n_j$, and the total 
absorption coefficients and the cross-sections are proportional to the 
imaginary part of $n_j$: $\xi_j=(2\omega/c){\rm Im}(n_j)$ and 
\begin{equation}
\label{sigopt}
\sigma_j=(2\omega/cN_e){\rm Im}(n_j), 
\end{equation}
respectively, where $N_e$ is the plasma number density.

For the normal modes description it is useful to introduce a complex 
variable $b$, or the two real parameters $q$ and $p$:
\begin{equation}
\label{b}
b\equiv q+{\rm i}p=\frac{n_L}{n_C}.  
\end{equation}
The polarization vectors in the rotating coordinates 
$e_{\pm}=2^{-1/2}(e_x\pm {\rm i}e_y)$, $e_0=e_z$, are then given by
\begin{eqnarray}
\label{evect}
e_{\pm}^j & = & \frac{1}{\sqrt{2}}C_j{\rm e}^{\mp{\rm 
i}\phi} ( K_j\cos\theta\pm 1),\\\nonumber
e_0^j & = & C_jK_j\sin\theta,
\end{eqnarray}
where
\[ K_j = b[1+(-1)^j\sqrt{1+b^{-2}}] \]
and $C_j=(1+|K_j|^2)^{-1/2}$ is the normalization. 

The polarization of the modes is determined by the complex quantity 
\begin{equation}
\label{alfa}
\alpha_j= |\alpha_j|\, {\rm e}^{2{\rm i}\chi_j},
\end{equation}
related to $b$ through
\begin{equation}
\label{alfab}
\alpha_j = -b^{-1}\mp\sqrt{1+b^{-2}}.
\end{equation}
The modulus of $\alpha_j$ determines the ellipticity
\begin{equation}
\label{ellipt}
{\cal P}_j = \frac{|\alpha_j| - 1}{|\alpha_j| + 1},
\end{equation}
with $|{\cal P}_j|$ being equal to the ratio                      
of the minor axis to the major axis of the polarization ellipse, and the 
sign of ${\cal P}_j$ determining the direction of rotation of the wave's 
electric field vector. The position angle $\chi_j$ is measured between the 
major axis of the polarization ellipse and the projection of the magnetic 
field $\bf B$ on the plane perpendicular to the wavevector $\bf k$. 

The Stokes parameters for a given normal mode can be written as
\begin{equation}
\label{stokes1}
I_j = I_j,\; Q_j = p_Q^jI_j,\; U_j = p_U^jI_j,\; V_j = p_V^jI_j,
\end{equation} 
where $I_j$ is the intensity of the mode, and $p_Q^j,p_U^j,p_V^j$ are 
defined by
\begin{equation}
\label{stokes2}
p_Q^j = \frac{2{\rm Re}(\alpha_j)}{|\alpha_j|^2 + 1},\;
p_U^j = \frac{2{\rm Im}(\alpha_j)}{|\alpha_j|^2 + 1},\;
p_V^j = \frac{|\alpha_j|^2 - 1}{|\alpha_j|^2 + 1}.
\end{equation} 
If the waves propagate independently, 
the total Stokes parameters are additive. Therefore,
for a radiation which can be regarded as a superposition of different waves 
with intensities $I_k$, they are
\begin{equation}
\label{stokes3}
I =\sum_k I_k,\; Q =\sum_k p_Q^kI_k,\; U =\sum_k p_U^kI_k,\; V = \sum_k 
p_V^kI_k.
\end{equation}
In the Monte Carlo approach this allows one 
to calculate the Stokes parameters of the outgoing radiation by summing up
the contributions of the individual photons.

\subsection{Normal modes in the system with cold plasma and vacuum}
\label{modes}
To find normal mode properties through 
equations (\ref{ni}-\ref{nc}) and (\ref{nj}) one must determine the 
permittivity tensor $\epsilon_{ab}$ and the magnetic permeability tensor 
$\mu_{ab}$. In the medium in which the plasma density is low, the 
contribution of the plasma to the dielectric tensor comprises a small 
departure from unity, so that $|\epsilon_{ab}^{(pl)}-1|\ll 1$. The magnetic 
permeability tensor is essentially $\mu_{ab}^{-1\,(pl)} = \delta_{ab}$.
Similarly, when the magnetic field is small, the vacuum contributions 
$|\epsilon_{ab}^{(vac)}-1|\ll 1$ and $|\mu_{ab}^{-1\,(vac)}-1|\ll 1$.
Therefore, for such a system the contributions of the plasma and the vacuum
can be added linearly, and the total dielectric tensor and total inverse 
magnetic permeability are
\begin{equation}
\label{tenstot}
\epsilon_{ab}=\epsilon_{ab}^{(pl)}+\epsilon_{ab}^{(vac)}-\delta_{ab},\;
\mu_{ab}^{-1}=\mu_{ab}^{-1\,(vac)}.
\end{equation}
Because of this, one can rewrite equations (\ref{ni}-\ref{nc}) in the 
form convenient for further discussion of normal mode properties
\begin{equation}
\label{nilc}
n_I=1+n_I^{pl}+n_I^{vac},\; n_L=n_L^{pl}+n_L^{vac},\; n_C=n_C^{pl},
\end{equation}
where the contributions to $n_I^{pl}, n_L^{pl}$ and  $n_C^{pl}$ come from
inserting $\epsilon_{ab}^{(pl)}-\delta_{ab}$ into equations 
(\ref{ni}-\ref{nc}) in place of $\epsilon_{ab}$ . 

Due to charge symmetry, the vacuum dielectric and magnetic permeability 
tensors $\epsilon_{ab}^{(vac)}$ and  $\mu_{ab}^{-1\,(vac)}$ are diagonal and 
real. Therefore, $n_C^{vac}\equiv 0$ (compare eq. \ref{nilc}), and 
from equations (\ref{nj}) and (\ref{nilc}) one can write for the refraction 
indices in vacuum 
\begin{eqnarray}
\label{njvac1}
n_1^{vac} & =& 1+n_I^{vac}-n_L^{vac}\\
\label{njvac2}
n_2^{vac} & =& 1+n_I^{vac}+n_L^{vac}.
\end{eqnarray}

The refraction indices $n_{1,2}^{vac}$ have been calculated for 
$\hbar\omega\ll m_ec^2$ and arbitrary magnetic field strength by 
\citet{tsai75}:
\begin{eqnarray}
\label{tsai1}
n_1^{vac} & =& 1+ \frac{\alpha_F}{4\pi}\eta_{\parallel}(h)\sin^2\theta\\
\label{tsai2}
n_2^{vac} & =& 1+ \frac{\alpha_F}{4\pi}\eta_{\perp}(h)\sin^2\theta,
\end{eqnarray}
where $\alpha_F$ is the fine-structure constant, $h= B_c/2B$, where
$B_c=m_e^2c^3/\hbar e$ is the critical field strength, and
\begin{eqnarray}
\label{etaperp}
\eta_{\perp}(h) & =& -4h\ln\Gamma (1+h)+4h^2\Psi (1+h) + 2h\ln h + 
2h[\ln(2\pi)-1] - 4h^2 + \frac{2}{3},\\\nonumber
\label{etapar}
\eta_{\parallel}(h) & =& 8\ln\Gamma_1 (1+h)-4h\ln\Gamma (1+h) 
-\frac{2}{3}\Psi (1+h) - 2h\ln h -2h^2 + 2h\ln(2\pi)\\
 & & + \frac{1}{3h} + \frac{1}{3} - 8L_1.
\end{eqnarray}
In these expressions, $\Psi(x)=d/dx \ln\Gamma (x)$,
\[ \ln\Gamma_1 (x)=\int_0^x dt \ln\Gamma(t) + 
\frac{x}{2}[x-1-\ln(2\pi)],\]
and $L_1= 1/3 + \int_0^1 dx \ln\Gamma_1 (1+x)\approx 0.249$.
Using equations (\ref{njvac1}-\ref{tsai2}) one 
can finally write
\begin{eqnarray}
\label{nIvac}
n_I^{vac} & = & \frac{1}{2}(n_1^{vac}+n_2^{vac})-1=\frac{\alpha_F}{8\pi}
\sin^2\theta [\eta_{\perp}(h)+\eta_{\parallel}(h)]\\
\label{nLvac}
n_L^{vac} & =& \frac{1}{2}(n_2^{vac}-n_1^{vac})=\frac{\alpha_F}{8\pi}
\sin^2\theta [\eta_{\perp}(h)-\eta_{\parallel}(h)].
\end{eqnarray}

The cold electron-proton plasma dielectric tensor is
\begin{equation}
\label{tenspl}
\epsilon_{ab}^{(pl)} = \delta_{ab} - \sum_{s=e,p} 
\left(\frac{\omega_{ps}^2}{\omega^2}\right)\Pi_{ab}^s,
\end{equation}
where the sum runs over electrons ($e$) and protons ($p$). In this 
expression, $\omega_{pe}=(4\pi N_e e^2/m_e)^{1/2}$ is the electron plasma 
frequency, $\omega_{pp}=\omega_{pe}(m_e/m_p)^{1/2}$ is the proton plasma 
frequency, and $\Pi_{ab}^s$ are the corresponding plasma polarization 
tensors. The general form of $\Pi_{ab}^s$, which takes into account both
the scattering processes and free-free absorption (but see \S \ref{scat})
is given in \S \ref{app_1}. For the case of pure scattering 
the polarization tensors are diagonal in the rotating coordinates and from
equations (\ref{app1}-\ref{app2}) we get 
\begin{equation}
\label{pialfa}
\Pi_{\alpha\alpha}^s = \frac{\omega}{\omega + \alpha\omega_{Bs}-{\rm 
i}\gamma_{rs}} = \frac{\omega}{\omega_{ts} + \alpha\omega_{Bs}},
\end{equation}
where $\alpha = -1,0,+1$, and $\gamma_{rs}$ is the radiative width: 
$\gamma_{re} = (2/3)(e^2/m_e c^3)\omega^2$ for electrons and 
$\gamma_{rp} =\gamma_{re}(m_e/m_p)$ for protons, and 
$\omega_{ts}=\omega+{\rm i}\gamma_{rs}$. Since in equation (\ref{pialfa}) 
the cyclotron frequency is defined as $\omega_{Bs}=(q_s B/m_s c)$, the 
electron cyclotron resonance occurs in $\Pi_{\alpha\alpha}^e$ for $\alpha = 
+1$, and the proton cyclotron resonance occurs in $\Pi_{\alpha\alpha}^p$ for 
$\alpha = -1$.

Inserting equations (\ref{tenspl}-\ref{pialfa}) into (\ref{ni}-\ref{nc}) and 
using equation (\ref{nilc}), we obtain
\begin{eqnarray}
\label{nIpl}
n_I^{pl} & = & - \frac{1}{4}\sum_{s=e,p}\frac{\omega_{ps}^2}{\omega} \left[ 
(1+\cos^2\theta) \frac{\omega_{ts}}{\omega_{ts}^{2} -\omega_{Bs}^2} + 
\sin^2\theta\frac{1}{\omega_{ts}}\right],  \\
\label{nLpl}
n_L^{pl} & = & -\frac{\sin^2\theta}{4}\sum_{s=e,p} 
\frac{\omega_{ps}^2}{\omega\omega_{ts}}\:\frac{\omega_{Bs}^2}{\omega_{ts}^
{2}-\omega_{Bs}^2},\\
\label{nCpl}
n_C^{pl} & = & \frac{\cos\theta}{2}\sum_{s=e,p} 
\frac{\omega_{ps}^2}{\omega}\: 
\frac{\omega_{Bs}}{\omega_{ts}^{2}-\omega_{Bs}^2}.
\end{eqnarray}

Damping of particle motion which enters the expressions 
(\ref{app1}-\ref{app2}) is usually very small, 
$\gamma_{\parallel, s}, \gamma_{\perp, s}\ll\omega$, and can be 
neglected in the derivation of normal mode polarization vectors. These can 
be thus obtained from equation (\ref{evect}) by taking expressions 
(\ref{nLvac}, \ref{nLpl}-\ref{nCpl}) and (\ref{nilc}) with 
$\gamma_{rs}\rightarrow 0$, and calculating parameter $b$ in equation 
(\ref{b}).

\subsection{Normal mode polarization properties}
\label{npol} 
As mentioned above, the normal mode polarization can be described by the 
complex quantity $b$, which determines the polarization characteristics 
through $\alpha_j$ (eqs \ref{alfa}-\ref{ellipt}). Calculating 
$n_L$ and $n_C$ in the manner described in \S \ref{modes} from the general 
expressions (\ref{app1}-\ref{app2}), and neglecting terms proportional to 
$\gamma_{\parallel, s}^2$ and $\gamma_{\perp, s}^2$, we obtain
\begin{eqnarray}
\label{q}
q & \approx & q_0 \left[\omega_{pe}^2\omega_{Be}^2(\omega_{Bp}^2-\omega^2) +
                  \omega_{pp}^2\omega_{Bp}^2(\omega_{Be}^2-\omega^2) +
		  \omega^2(\omega_{Be}^2-\omega^2)
		   (\omega_{Bp}^2-\omega^2)V(h)\right],\\
\nonumber
p & \approx & \frac{q_0}{\omega}[                   
  \gamma_{\parallel,e}\omega_{pe}^2\omega_{Be}^2(\omega^2-\omega_{Bp}^2) +	
  \gamma_{\parallel,p}\omega_{pp}^2\omega_{Bp}^2(\omega^2-\omega_{Be}^2) \\
\label{p}  
 &  &  \;\;\;\;\; + 2\omega^4(\gamma_{\perp,e}(\omega^2-\omega_{Bp}^2) +
            \gamma_{\perp,p}(\omega^2-\omega_{Be}^2))V(h)],
\end{eqnarray}
where we denoted
\[ q_0 = \frac{\sin^2\theta}{2\cos\theta}
       \left[\omega_{pe}^2\omega_{Be}\omega(\omega^2-\omega_{Bp}^2) +
       \omega_{pp}^2\omega_{Bp}\omega(\omega^2-\omega_{Be}^2)\right]^{-1} \]
and
\[ V(h)=\frac{\alpha_F}{2\pi}
                   [\eta_{\perp}(h)-\eta_{\parallel}(h)].\]
For most of frequencies and angles $q^2\gg p^2$ and $\alpha_j$, ${\cal P}_j$ 
and $\chi_j$ can be approximated as follows 
\citep{pav79,1980Ap&SS..73...33P}:\begin{eqnarray}
\label{alfap1}
\alpha_j & = & \frac{1\pm \sqrt{1+q^2}}{q}\left[1\mp\frac{{\rm i}p}
               {q\sqrt{1+q^2}}\right],\\
\label{pj1}
{\cal P}_j & = & \pm \frac{1}{|q|+ \sqrt{1+q^2}},\\
\label{chi1}
\chi_j  & = & \mp \frac{p}{2q \sqrt{1+q^2}} + 
                  \frac{\pi}{2}\frac{1\pm {\rm sign}\,q}{2} {\rm sign}\,p.
\end{eqnarray}
Therefore, the modes are nearly orthogonal ($\chi_1\simeq 0, 
\chi_2\simeq\pm\pi/2$ and ${\cal P}_1 = - {\cal P}_2$), and when $|q|\ll 1$, 
they are circularly polarized, $|{\cal P}_1| = |{\cal P}_2|\simeq 1$, and 
they are linearly polarized, $|{\cal P}_1| = |{\cal P}_2|\simeq 0 $ for 
$|q|\gg 1$. The latter case is realized for the electron-proton plasma at 
photon energies far from the critical points and in 
the range $\omega\ll\omega_{Be}$ of interest in this paper. 

The critical points occur at photon frequencies for which $q(\omega)=0$. 
From equation (\ref{q}) one can see that for the system considered here 
there exist three such points. The first two points are located 
very close to the electron and the proton cyclotron frequencies, and the 
location of the third one, called the vacuum resonance frequency, is 
approximately
\begin{equation}
\label{omv}
\omega_V\approx\omega_{pe}\left(\frac{\alpha_F}{2\pi}
                   [\eta_{\parallel}(h)-\eta_{\perp}(h)]\right)^{-1/2}.
\end{equation}
The vacuum resonance arises when the contributions of plasma and  vacuum 
to the dielectric tensor become comparable to each other. Because $\omega_V$
depends on plasma density, the actual frequency at which the resonance 
occurs varies in the inhomogeneous atmosphere of the neutron star. 

The normal mode polarization properties near the critical points are 
determined by the parameter $p$, which depends on the mechanism of 
absorption through the corresponding damping rates (see eq. \ref{p}). 
Since near these points $q^2\ll p^2$, equations (\ref{alfap1}-\ref{chi1}) 
become invalid and must be replaced by other approximate expressions.
Thus, for $\alpha_j$ we have 
\begin{equation}
\label{alfap2}
\alpha_j =  [1\pm \sqrt{1-p^2}]/{\rm i} p,
\end{equation}
and when $|p|\leq 1$ 
\begin{equation}
\label{pj2}
{\cal P}_j = \pm \frac{\sqrt{1-p^2}}{1+|p|},\; 
\chi_j = \frac{\pi}{4} {\rm sign}\,p,
\end{equation}
hence the modes are circular but the polarization ellipses coincide, and 
when $|p|\geq 1$
\begin{equation}
\label{pj3}
{\cal P}_j =0,\;
\chi_j  = \frac{\pi}{4}({\rm sign}\,p \pm 1)\mp 
          \frac{1}{2}\tan^{-1}\sqrt{p^2-1},
\end{equation}            
and the modes are linear but the polarization ellipses neither coincide nor 
are orthogonal. In the limiting case where $q=0$ and $|p|=1$, the normal 
modes are completely linear, ${\cal P}_j =0$, and entirely coincide, 
$\chi_1=\chi_2=\pm \pi/4$, i.e. they are completely nonorthogonal.    

For each of the critical frequencies, the point where the modes are 
completely nonorthogonal (also called a "mode collapse point") occurs 
at the critical angle $\theta_c$ such that $|p(\omega_c, \theta_c)|=1$.
Therefore, as evident from equations (\ref{pj2}-\ref{pj3}), the behavior of 
ellipticities and position angles near the critical points $\omega_c$ 
depends on the relationship between $\theta_c$ and the photon propagation 
angle $\theta$. At the passage through $q=0$ points there is either 
a change in $\chi_j$ by $\pm \pi/2$, when $\theta< \theta_c$ ($|p|\leq 1$), 
or the change in the sign of ${\cal P}_j$, if $\theta> \theta_c$ ($|p|\geq 
1$). 
This means that the separation of the modes into an ordinary and 
an extraordinary one in the presence of critical points becomes ambiguous
\citep{pav79,1980Ap&SS..73...33P,bul96}.

\subsection{Scattering and absorption cross-sections}
\label{scat}
The differential cross-section for the scattering on electrons 
in a hot magnetized plasma is given by
\begin{equation}
\label{compt}
\frac{d^2\sigma^e}{d\omega'd\Omega'}(\omega\theta\rightarrow\omega'\theta')= 
r_0^2 \frac{\omega'}{\omega}\int dp f(p)|\langle 
e|\Pi|e'\rangle|^2\delta(\omega+\Delta\omega -\omega').
\end{equation}
Here $r_0$ is the classical electron radius, 
$\omega,\theta,e$ and $\omega',\theta',e'$ are the photon energy, the angle 
with respect to the magnetic field $\bf B$, and polarization vector before 
and after scattering, respectively, and $\Delta\omega=\omega'-\omega$. The 
electron momentum distribution $f(p)$ is taken to be a one-dimensional 
Maxwellian for electrons in the lowest Landau level. Averaging equation 
(\ref{compt}) over $\phi'$ we obtain
\begin{equation}\label{comptav}
\langle\frac{d^2\sigma^e}{d\omega'd\Omega'}\rangle_{\phi'}
(\omega\theta\rightarrow\omega'\theta') = r_0^2 \frac{\omega'}{\omega}\int 
dp f(p)\sum_{\alpha}|\Pi_{\alpha\alpha}^{e(r)}|^2 |e_{\alpha}|^2 
|e'_{\alpha}|^2\,\delta(\omega+\Delta\omega -\omega').
\end{equation}
The components of the electron polarization tensor, with 
first-order relativistic corrections in the electron $p$ and photon $k,k'$ 
momenta are
\begin{eqnarray}
\label{pi+}
\Pi^{e(r)}_{++} & = & 1- \frac{\omega_{Be}}{\omega'+\omega_{Be}-pk'+k'^2/2 
- {\rm i}\gamma_{re}},\\
\label{pi-}
\Pi^{e(r)}_{--} & = & 1+ \frac{\omega_{Be}}{\omega-\omega_{Be}-pk-k^2/2 - 
{\rm i}\gamma_{re}},\\
\label{pi0}
\Pi^{e(r)}_{00} & = & 1+ \frac{(p+k/2)(p+k-k'/2)}{\omega-pk-k^2/2 - {\rm 
i}\gamma_{re}}- \frac{(p-k'/2)(p-k'+k/2)}{\omega'-pk'+k'^2/2 - {\rm 
i}\gamma_{re}}.
\end{eqnarray}
For coherent scattering, i.e. when $\Delta\omega=0$, these expressions 
reduce to those given for a cold plasma by equation (\ref{pialfa}). In that 
case, after integrating (\ref{compt}) over the final states, we obtain the 
scattering cross-sections from one mode to another
\begin{equation}
\label{sigije}
\sigma^e_{ij}=\sigma_{T}\sum_{\alpha} |e^i_{\alpha}|^2 
|\Pi_{\alpha\alpha}^{e}|^2 A^j_{\alpha},
\end{equation}
where
\[ A^j_{\alpha} = \frac{3}{4}\int d\cos\theta |e^j_{\alpha}|^2, \]
and $\sigma_{T}$ is the Thomson cross-section.
In the transverse mode approximation, the polarization modes satisfy the 
completeness property $\sum_{j=1,2} |e^j_{\pm}|^2 = (1+\cos^2\theta)/2$ and
$\sum_{j=1,2} |e^j_{z}|^2 = \sin^2\theta$, so that $\sum_{j=1,2} 
A^j_{\pm} = 1$ and  $\sum_{j=1,2} A^j_{z} = 1$. Using these relations, the 
total electron scattering cross-section 
$\sigma^e_i=\sigma^e_{ii} + \sigma^e_{ij}$ in the cold plasma limit is
\begin{equation}
\label{sigie}
\sigma^e_{i}=\sigma_{T}\sum_{\alpha} |e^i_{\alpha}|^2 
|\Pi_{\alpha\alpha}^{e}|^2.
\end{equation}

In the ionized plasma studied in this paper, protons also contribute to 
the photon scattering. However, under condition of thermal equilibrium, 
proton mean thermal velocities are $\sim (m_p/m_e)^{1/2}$ times smaller then 
the electron ones, and for the considered $kT=10$ keV plasma Comptonization 
effects, as of equations (\ref{compt}-\ref{pi0}), are negligible. Thus, 
analogously to equations (\ref{sigije}) and  (\ref{sigie}), we have for the 
proton scattering cross-sections
\begin{equation}
\label{sigijp}
\sigma^p_{ij}=\left(\frac{m_e}{m_p}\right)^2 \sigma_{T}\sum_{\alpha} 
|e^i_{\alpha}|^2 |\Pi_{\alpha\alpha}^{p}|^2 A^j_{\alpha},
\end{equation}
and
\begin{equation}
\label{sigip}
\sigma^p_{i}= \left(\frac{m_e}{m_p}\right)^2 \sigma_{T}\sum_{\alpha} 
|e^i_{\alpha}|^2 |\Pi_{\alpha\alpha}^{p}|^2.
\end{equation}

As described by equations (\ref{sigije}) and (\ref{sigijp}), a photon can 
switch between the polarization modes in a scattering act. The mode 
switching probability for the two modes is defined
\begin{equation}
\label{switch}
p^s_{ij}=\frac{\sigma^s_{ij}}{\sigma^s_{ij}+\sigma^s_{ii}},
\end{equation}
where again the subscript ($s$) is for electrons ($e$) or protons ($p$).

The total scattering cross-section is the sum of the electron and proton 
components, namely
\begin{equation}
\label{sigtot}
\sigma_{i}=\sigma^e_{i}+\sigma^p_{i}.
\end{equation}
This expression is essentially equivalent to that given through the optical 
theorem by equation (\ref{sigopt}), when $\gamma_{\parallel, s} = 
\gamma_{\perp , s} = \gamma_{rs}$ (see \S \ref{app_1}). However, it is 
important to note that when vacuum polarization effects are considered, 
definition of the modes  through the optical theorem becomes ambiguous. 
Below the vacuum frequency $\omega_{V}$, the extraordinary mode is 
denoted by $(j=1)$ and the ordinary one by $(j=2)$ in equation 
(\ref{sigopt}), whereas this notation is reverted above $\omega_{V}$.
Thus one has to be careful to choose the correct branches
of complex roots so that the opacities found using the optical theorem 
(eq. \ref{sigopt}) match these calculated directly using the 
polarization vectors. The extraordinary mode is also referred to as the 
low-cross-section mode.

Free-free absorption is also a significant source of the opacity over 
the frequencies and densities of interest. The cross-section for this 
process can be found through the optical theorem by taking 
$\gamma_{\parallel, s} = \gamma_{\parallel, cs}$, 
$\gamma_{\perp, s} = \gamma_{\perp, cs}$ in equations 
(\ref{app1}-\ref{app2}). However, as has been recently obtained by 
\citet{pot03}, these expressions are inaccurate at $\omega\la \omega_{Bp}$.
To derive the correct expressions for the 
absorption cross-section one has to take  
into account the effects of the finite proton mass on the 
process, which allow the absorption to occur in proton-proton collisions and 
also lead to the modifications of the absorption in electron-proton 
collisions due to the proton motion. For $kT=10$ keV, the contribution to 
the absorption from proton-proton collisions is negligible and we can write 
for the free-free absorption cross-section
\begin{equation}
\label{abs}
\sigma^{ff}_{i}=  \sigma_{a}\sum_{\alpha}
\frac{\omega^4}{(\omega + \alpha\omega_{Be})^2 (\omega+\alpha\omega_{Bp})^2
+ \omega^2\gamma^2_{\alpha}} |e^i_{\alpha}|^2 g_{\alpha},
\end{equation} 
where the absorption coefficient
\begin{equation}
\label{siga}
 \sigma_{a}=4\pi^2\alpha_F^3\frac{\hbar^2c^2}{m_e\omega^3}
\frac{N_i}{(\pi m_e kT/2)^{1/2}} (1-{\rm e}^{-\hbar\omega/kT}) 
\end{equation}
is corrected for stimulated emission, and the dimensionless magnetic Gaunt 
factors, $g_{\pm}=g_{\perp}$ and $g_{0}=g_{\parallel}$, are evaluated 
according to the formulas given by \citet{mesz92}. The effective damping 
in equation (\ref{abs}) is included through $\gamma_{\alpha} 
=\gamma_{\alpha, ce} + \gamma_{re}+\gamma_{rp}$, where the collisional 
damping 
$\gamma_{\alpha, ce}=(\sigma_{a}g_{\alpha}/\sigma_{T})\gamma_{re}$.

\begin{figure}
\plotone{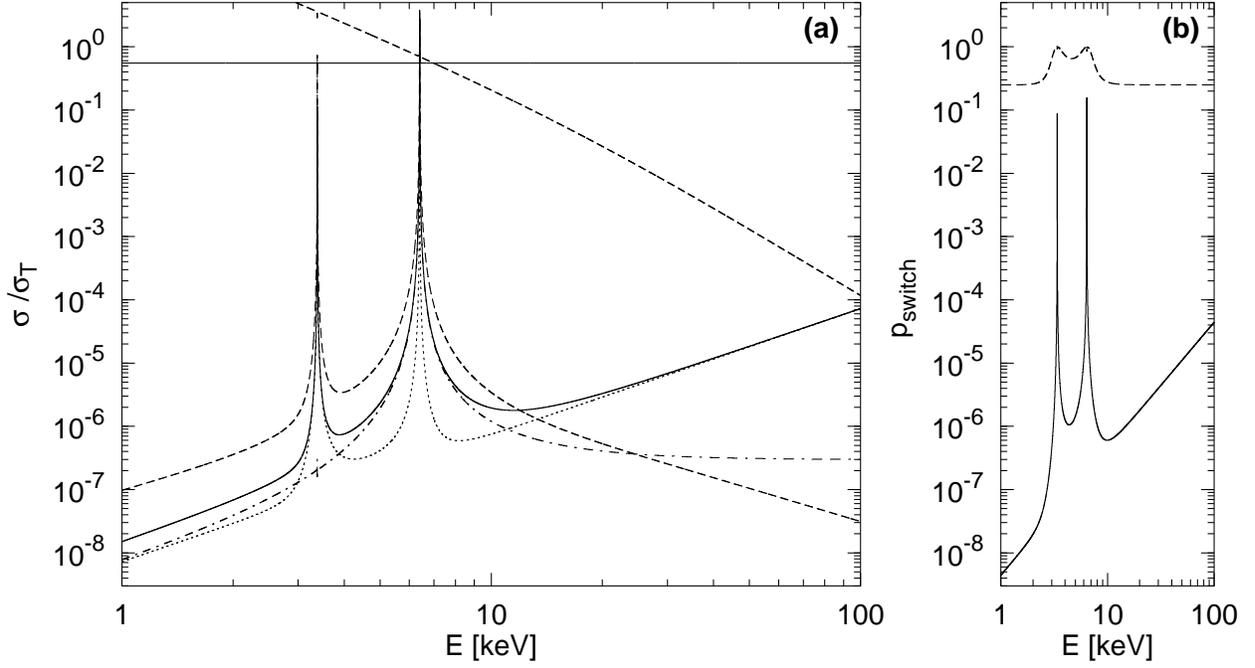}
\caption{(a) Scattering and absorption cross-sections as a function of 
energy at the density $\rho=\rho_{0,1}\simeq 220$ g\,cm$^{-3}$
(corresponding to $z_{0,1}=10H$) and propagation angle  
$\theta=48^\circ$ ($\cos\theta=0.67$).
The solid lines show the total (electron + proton)
scattering cross-sections for  
both the high- and the low-cross-section mode. 
The individual contributions to the total cross-section in the 
extraordinary mode from electron and proton scattering are shown by 
dotted and dash-dotted lines, respectively. Note that the vacuum resonance
at $\hbar\omega_{V}\simeq 3.38$\,keV only slightly affects the proton 
scattering cross-section. The absorption cross-sections for the two 
polarization modes are shown by the dashed lines. 
(b) The mode switching 
probabilities: low to high mode (dashed line) and high to low 
mode (solid line).
\label{cross}}
\end{figure}

We present the  scattering and absorption cross-sections as a function of 
photon energy in Figure \ref{cross}a. Vacuum polarization affects 
the normal mode polarization vectors, which leads to redistribution of the
total scattering cross-section between the modes: at the vacuum resonance 
frequency $\omega_V$, $\sigma_1\approx\sigma_2\approx\sigma_T$. The 
absorption cross-section is also enhanced at $\omega_V$ in the 
low-cross-section mode 
(for $B=B_{15}$ and density $\rho=\rho_{0,1}\simeq 220$ g\,cm$^{-3}$
(see below), $\hbar\omega_V\simeq 3.38$\,keV in Fig. \ref{cross}a). 
The vacuum 
resonance only slightly affects the proton scattering cross-section for the 
density presented. At this density, the proton scattering cross-sections for 
the two polarization modes are of similar magnitude at the 
vacuum frequency. The opacity in the extraordinary mode
below $10$\,keV is dominated by absorption. We note that in the wings
of the proton resonance ($\hbar\omega_{Bp}\simeq 6.4$\,keV)
the proton scattering cross-section dominates over the 
electron scattering. The electron scattering is resonant 
at the proton cyclotron frequency due to influence of the 
proton resonance on the normal mode polarization vectors.
Figure \ref{cross}b shows the mode switching probability
as a function of energy for the same conditions. The 
low- to high-cross-section mode switching probability is greatly enhanced
around the resonances, while far from them it settles at the
value of $1/4$.

\section{MONTE CARLO RADIATIVE TRANSFER SOLUTION}
\label{mc}
To calculate the spectra and polarization properties 
of radiation emerging from magnetized neutron star atmosphere we use a Monte 
Carlo method based on 
\citet{bul97}. 
Assuming an isothermal hydrogen atmosphere with $kT=10$\,keV and
the density profile of the form $\rho=\rho_0 {\rm e}^{z/H}$,
we use an adaptive mesh  to produce a table 
of physical depth $z$ as a function of optical depth $\tau$ for the 
low-cross-section mode, for
various values of model parameters: magnetic field strength
and inclination and the scaleheight $H$.
The normal modes polarization vectors are derived in the cold plasma 
approximation, and the scattering and absorption cross-sections are 
obtained from equations (\ref{sigtot}) and (\ref{abs}). 
The thermal effects modify the modes for 
photon frequencies only very close to $\omega_{Bp}$, and their net result 
on the scattering cross-sections is to broaden the proton cyclotron 
resonance \citep[see, e.g,][]{mesz92}. Therefore, these effects, although 
important for the scattering process, can be neglected in the optical depth 
derivations on the grid.
The optical depth 
depends on the photon energy and the propagation angle $\theta$, and in the 
case of a tilted magnetic field on the azimuthal angle $\phi$ as well.  
The changes in the physical depth can be very sharp when approaching the
vacuum resonance, and thus the mesh must be carefully chosen. We use 260 
logarithmically spaced zones in photon frequency from $\hbar\omega = 
10^{-2.6}m_ec^2$ to $\hbar\omega = m_ec^2$, 90 linearly spaced zones in the 
angle $\theta$ in the range $[0, \pi/2]$ in modeling the case with 
perpendicular magnetic field (for the model with a tilted field we use 
120 linearly spaced zones in $\theta$ and $\phi$ in the range $[0, \pi]$), 
and 276 logarithmically spaced zones in optical depths from $10^{-2}$ to 
$10^{4}$, which is well sufficient to achieve the accuracy required.

The transfer of radiation through the neutron star atmosphere is solved
with the method of Monte Carlo simulations. Photons  are injected 
at a single layer deep in the atmosphere. 
The energies of the photons
are selected from a 10-keV blackbody (Planck) distribution, and their 
propagation directions are chosen randomly from a uniform distributions in 
$\cos\theta$ and $\phi$. At each interaction (scattering or absorption)
we generate an optical depth traveled by the photon, and from the table of 
physical depth versus optical depth for given $\omega$, $\theta$ and 
$\phi$, we obtain the new physical depth of the photon. If the photon 
escapes from the atmosphere, its energy and propagation direction are kept, 
together with its polarization characteristics, which are calculated in 
terms of the Stokes parameters (eqs (\ref{alfa}-\ref{alfab}) and
(\ref{stokes1}-\ref{stokes2}); $b$ in equation (\ref{alfab}) being derived 
at the depth of the last interaction). Otherwise, from equations 
(\ref{sigie}), (\ref{sigip}) and (\ref{abs}), we calculate the 
cross-sections and determine which interaction process (electron scattering, 
proton scattering or absorption) occurs. When the electron scattering is 
chosen, we use the differential cross-section of equation (\ref{comptav}) 
to select the new energy and the angle of the photon, and determine whether 
it switches polarization modes. 
In our procedure we do not integrate the cross-sections over the electron 
momentum $p$
in equation~(\ref{comptav}), but draw the value of $p$ from 
the thermal distribution.
For the proton scattering the 
new photon parameters are selected using equation (\ref{sigijp}). Finally, 
in the case of absorption, we draw the new energy of the photon from the 
thermal distribution and select in random the new propagation angles. Then 
the process is repeated until the photon escapes from the atmosphere. 
 We impose 
a reflection-like condition at the bottom of the atmosphere:
a photon that wanders below the production depth
is lost and a new one is drawn  from a thermal distribution in 
its place. In a typical simulation we transfer  between
$10^9$ and $10^{10}$ photons, a significant increase in comparison
with $10^5$ in the simulations of  \citet{bul97}.

In the present simulations,
the effects of mode switching are accounted for in the same manner as in the 
work of \citet{bul97} --- the photon is absorbed when switching to the 
high-cross-section mode. For higher photon energies, electron scattering 
dominates the opacity in the high-cross-section mode and the probability 
$p_{h\rightarrow l}$ of switching to the low-cross-section mode is very low 
(see Figure~\ref{cross}b).
Thus, the photon may require many scatterings until it switches its mode 
again, during which it may substantially change its energy and is likely to 
be absorbed. For lower photon energies, $\hbar\omega\la 10$ keV, and 
in the range of densities of interest, absorption is the main source of 
opacity for the high-cross-section mode everywhere, except at the proton 
cyclotron resonance. For scattering with the mode switching in either vacuum 
or proton resonance, the probability of returning to 
the low-cross-section mode in the next scattering is high. However, in the 
case of electron scattering, the fractional energy change in the process is 
much greater than the widths of the resonances and the photon will be 
usually Comptonized out of the resonance, where $p_{h\rightarrow l}$ is  
low. Thus, the applied procedure may not be accurate only for the proton 
scattering at $\omega\simeq\omega_{Bp}$, where the process is already 
approximately described because of the thermal effects neglection in the
derivation of the optical depths. Likewise in \citet{bul97}, the physical 
depth of the photon is not changed at mode switching, because the mean 
free path is small in the high-cross-section mode and the net distance 
traveled by the photon in this mode is negligible.

\section{RESULTS}
\label{results}  
We present our numerical results for the case of the fully ionized hydrogen 
atmosphere with the magnetic field strength $B=10^{15}$ G. This value is 
chosen in order to conveniently discuss the proton effects on the spectral 
and polarization characteristics of soft gamma-ray repeaters. The proton 
cyclotron resonance thus occurs at $\hbar\omega_{Bp}= 6.4$ keV, and the 
departures from the transversality of the normal modes at $\hbar\omega\la 2$ 
keV and at the highest plasma densities considered, do not influence the 
results.

Since, as shown in \citet{bul97}, the shape of the emergent  spectrum 
highly depends on the photon  production depth, in the following we 
compare the results obtained for two injection  depths (in the units of the 
scaleheight): $z_{0,1}=10H$ and $z_{0,2}=13H$, which translates into the 
plasma density of $\rho_{0,1}\simeq 220$ g\,cm$^{-3}$ and $\rho_{0,2}\simeq 
4.4\cdot 10^3$ g\,cm$^{-3}$, respectively, for the assumed density at the 
surface $\rho_0=10^{-2}$ g\,cm$^{-3}$ (corresponding to the number density 
$N_e\simeq 6\cdot 10^{21}$ cm$^{-3}$). 
We consider two cases of the magnetic field inclination: the field
perpendicular to the surface and the field tilted at an angle of $45^\circ$ 
to the surface.

\begin{figure}
\plotone{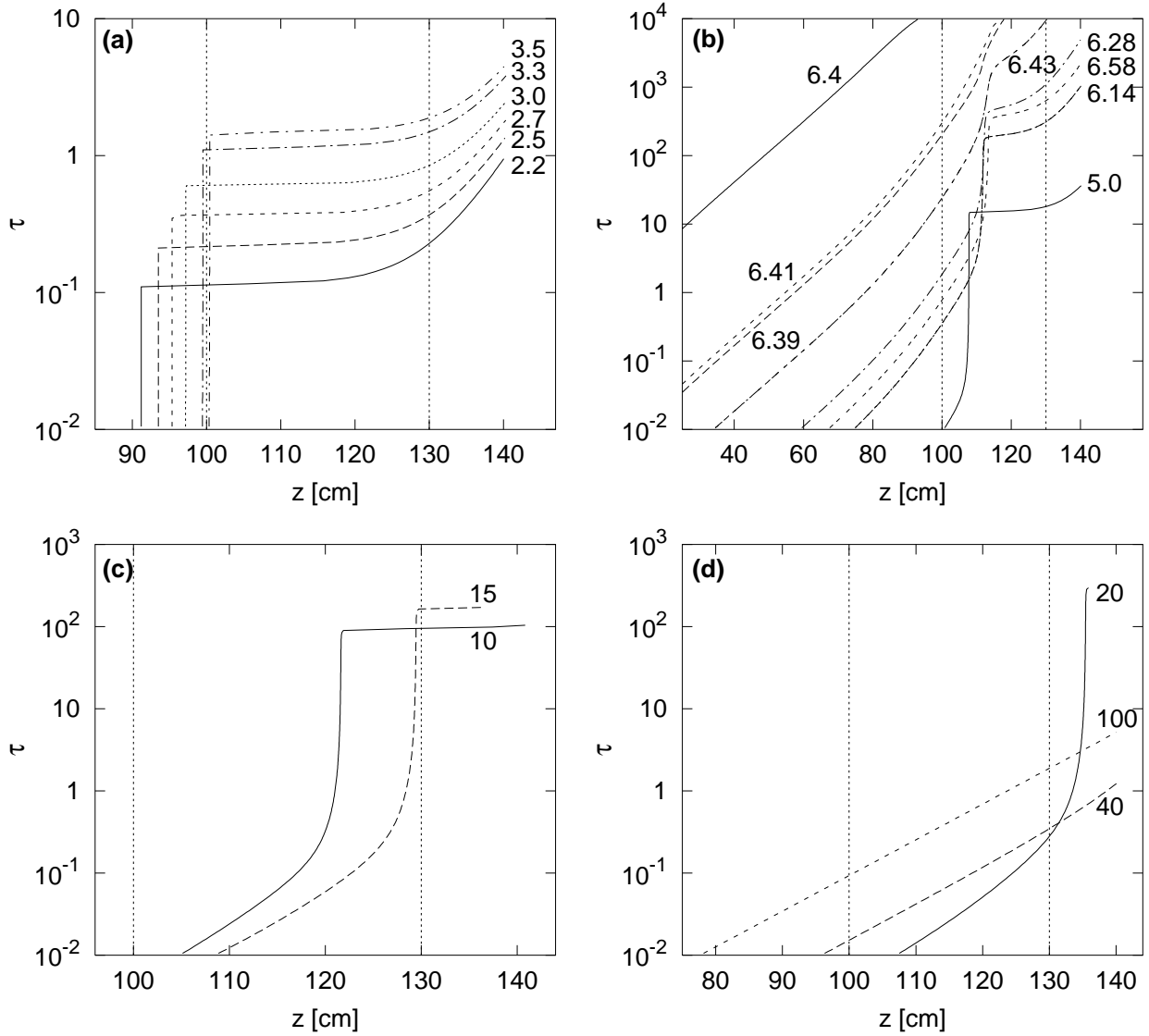}
\caption{The dependence of the optical depth on the physical depth
in the atmosphere with $H=10$\,cm and the magnetic field perpendicular to 
the surface, for the propagation angle of $48^\circ$. The vertical lines 
denote the photon injection depths in our model simulations, and the numbers 
next to the curves indicate photon energies in keV. Panel (a) shows the 
region of photon energies around $\hbar\omega_{V,1}$, while (c) illustrates 
the range around $\hbar\omega_{V,2}$. The energy range  around the proton 
resonance $\hbar\omega_{Bp}=6.4$\,keV, where the vacuum resonance 
disappears, is shown in (b). Panel (d) illustrates the energy region above 
the resonances.\label{tauz}}
\end{figure}

Figure \ref{tauz} shows the example behavior of 
optical depth $\tau$ as a function of physical depth $z$ for the 
low-cross-section mode at various photon energies and the propagation 
direction $\theta\approx 48^o$ ($\cos\theta=0.67$), for the atmosphere
model with scaleheight $H=10\,$cm, and 
the magnetic field  normal to the 
neutron star surface.
 In this plot, the vacuum resonance manifests itself as 
a sharp increase in the optical depth occurring at the physical depth 
corresponding to the resonant density, at which scattering and absorption 
cross-sections are greatly enhanced at the given photon frequency (see \S 
\ref{scat}). The photon energy corresponding to the location of the resonant 
physical depth at $z_{res}=z_{0,1}$ equals $\hbar\omega_{V,1} = 3.38$ keV 
(Fig. \ref{tauz}a). For $\omega > \omega_{V,1}$ the resonant depth 
increases, reaching $z_{res}=z_{0,2}$ at $\hbar\omega_{V,2} = 15.15$ keV 
(Fig. \ref{tauz}c). The peculiar behavior occurs at energies near the proton 
cyclotron resonance. At physical depths for which $\omega_V$ of equation 
(\ref{omv}) is approximately equal to $\omega_{Bp}$, the plasma 
contribution to the dielectric tensor is much greater than that of vacuum 
and the vacuum resonance does not exist, as can be seen in Figure 
\ref{tauz}b. Because proton scattering and absorption cross-sections are 
large near $\omega_{Bp}$, the optical depths are considerably enhanced in 
this energy range. The resonant depths for photon energies shown in Figure 
\ref{tauz}d lie far below $z_{0,2}$.

\subsection{Spectra}
\label{spectra}
\begin{figure}
\plotone{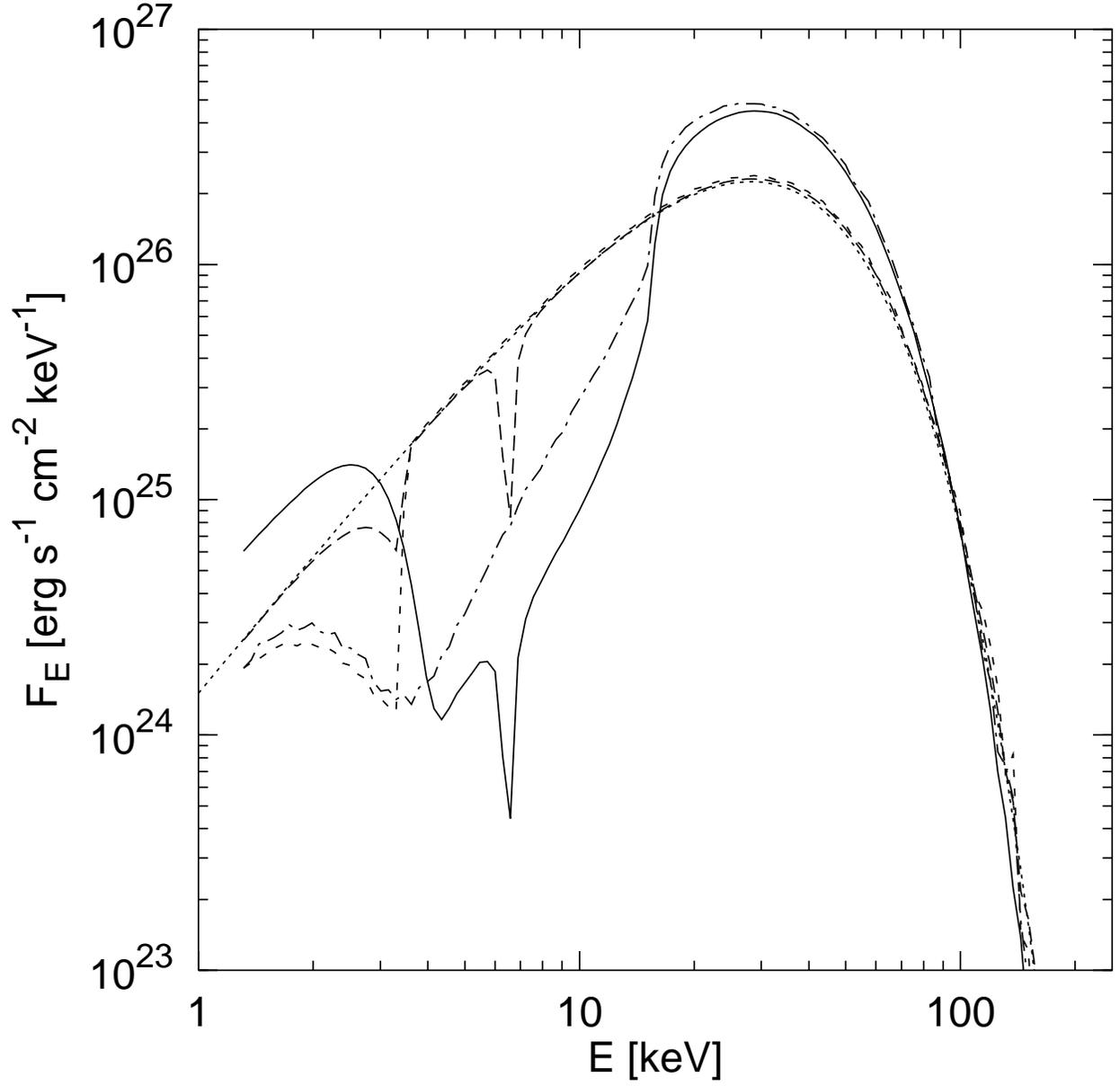}
\caption{Model spectra at the top of the atmosphere for the case of the 
magnetic field perpendicular to the surface and the scaleheight $H=10$\,cm. 
The dotted line is the input Planck spectrum with $kT=10$\,keV.
The solid line represents the case of $z=z_{0,2}$ injection, while the 
long-dashed line corresponds to the case of  $z=z_{0,1}$. The spectra for 
the model without protons are also presented 
for comparison: dot-dashed line is for $z=z_{0,2}$, and short-dashed line 
for $z=z_{0,1}$. 
\label{s10ep}}
\end{figure}

Figure \ref{s10ep} shows photon spectra at the top of the atmosphere 
for the model with 
scaleheight $H=10$\,cm and the magnetic field perpendicular to the surface 
($B=10^{15}$\,G). When the radiation is produced very deep 
in the atmosphere ($z=z_{0,2}$), the absorption-like feature resulting 
from the vacuum polarization effects is distinct and broad, and the upper 
bound on the feature is at $\hbar\omega_{V2}\sim 15$\,keV. 
At energies below and above those 
at which the absorption-like feature forms, the spectrum shows a significant 
excess over the blackbody flux. For the smaller production depth 
$z=z_{0,1}$, the overall spectrum is planckian and the absorption-like 
feature is weakly marked and bound in a narrow energy range near
 $\hbar\omega_{V1}\sim 3$\,keV. At either production depth, 
 the proton cyclotron line feature at 
$\omega\approx\omega_{Bp}$ is clearly visible in the spectra.

The absorption-like features in the spectra owe their existence to the 
influence of vacuum polarization on the absorption and scattering 
cross-sections, and in the latter case the effect of vacuum on the mode 
switching probability is most important. 
Optical depths for photons interacting in the resonance are large, thus 
their propagation through the atmosphere requires several 
absorption or scattering acts. Absorption can set the photon energy 
out of the resonance, to the region of small optical depths, thus leading to 
its escape from the atmosphere and creating a relative deficit at the 
original energy. The net effect of scattering is similar --- 
at the resonance the mode switching probability is close to unity (see 
Fig.~\ref{cross}b), which results in photon absorption at almost every 
scattering act (see \S \ref{mc}). In consequence, the effect of 
Comptonization in shaping the spectrum is negligible in the energy range 
where the absorption-like feature occurs \citep[cf.][]{bul97}.

\begin{figure}
\plotone{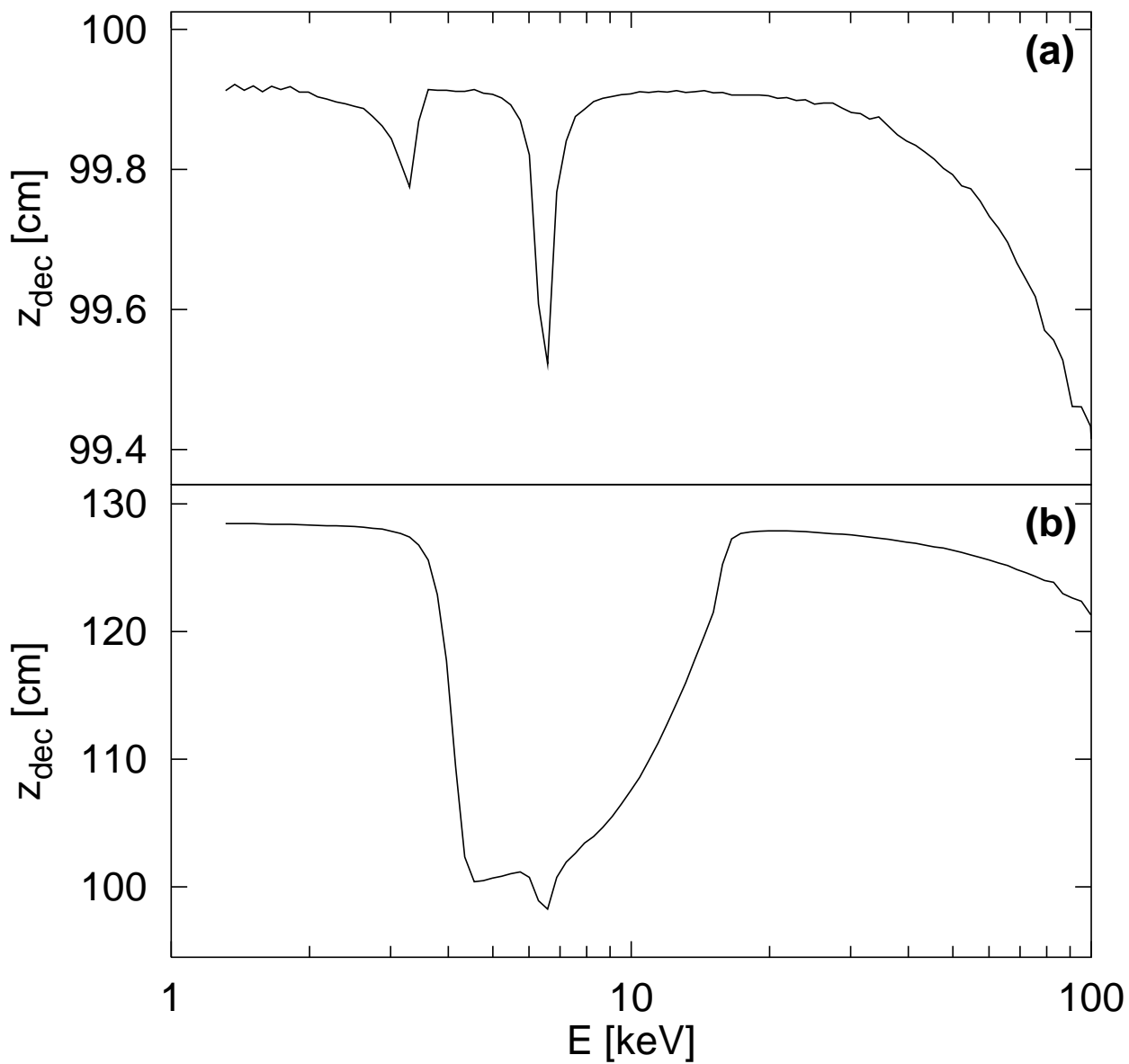}
\caption{The mean decoupling depths for the photon propagation angle of 
$48^\circ$ and for the two production depths we consider: $z=z_{0,1}$ (a)
and $z=z_{0,2}$ (b) ($H=10$\,cm). Note 
the difference in vertical scales. In the case of the $z=z_{0,2}$ injection, 
the vacuum and the proton resonances blend each other.
\label{zmean}}
\end{figure}

The upper bounds on the
absorption-like features in the spectra shown in Figure \ref{s10ep} 
correspond roughly to the frequencies $\omega_{V,1}$ and  $\omega_{V,2}$, at 
which resonant physical depths are  equal to the production depths. As can 
be seen in Figure \ref{tauz}, at energies above $\hbar\omega_{V,1}$ the 
optical depths for photons produced at the smaller production depth 
$z_{0,1}=10H$ are small. Thus most of them are generated above decoupling 
layer (i.e. at $z<z_{dec}$) and escape freely from the atmosphere. The 
location of the decoupling layer, estimated roughly as a mean physical depth 
of the last interaction, is shown in Figure \ref{zmean}, for the 
representable value of the propagation angle $\theta=48^\circ$. For energies 
at which absorption-like feature forms, injection depth is on average above 
the mean decoupling depth, but the differences are small. This and the fact 
that the feature is shallow in this case results from the optical depths of 
photons which form the vacuum feature being only slightly enhanced 
compared to the regions outside of it, so that their propagation is only 
weakly influenced by interaction processes. The situation is different for 
photons injected at larger depth $z_{0,2}=13H$, because in this case the 
optical depths of photons forming the feature are very large. As a result, 
absorption and scattering processes modify the spectrum significantly, 
reducing the photon flux in the feature and redistributing it to the regions 
of smaller optical depths, thus forming ``bumps'' in the spectrum at 
energies below and above the vacuum feature. Correspondingly, the mean 
decoupling depths in the feature (Fig. \ref{zmean}b) are also considerably 
reduced.   Note also that the high-energy part of the spectrum 
($\omega\gg\omega_{Bp}$) is steeper then the blackbody because the 
cross-section for scattering, that is dominant in this energy range,  
increases as $\omega^{2}$, which shifts the distribution of escaping photons 
to lower energies \citep[see also Figs \ref{tauz}d and \ref{zmean}]{mil96}.  

 The absorption line at the proton cyclotron 
frequency $\hbar\omega_{Bp}=6.4$ keV is prominent in the spectra. The line 
is distinct and broader than the natural or thermal width, as discussed in, 
e.g., \citet{2001MNRAS.327.1081H}. The strength of the line is greater for 
the larger photon production depth $z_{0,2}$, which can be traced back to 
the differences in opacities at $z_{0,1}$ and $z_{0,2}$ at 
$\omega\approx\omega_{Bp}$ (Fig. \ref{tauz}b). The decoupling density is 
also reduced at the proton resonance and is lower than the decoupling 
densities in the absorption-like feature (Fig. \ref{zmean}). This results 
from enhanced absorption and the dominance of proton scattering over Compton 
electron scattering near~$\omega_{Bp}$.

Effects of protons on the plasma response properties are revealed  not only 
as the absorption line feature in the spectrum, but also affect its overall
shape \citep[see, 
e.g.,][]{2001MNRAS.327.1081H,2003MNRAS.338..233H,2003ApJ...583..402O,zan01}. 
This is exemplified in Figure \ref{s10ep} by short-dashed and dot-dashed 
lines, which show the spectra for the atmosphere model without protons, for 
the production depth $z_{0,1}$ and $z_{0,2}$, respectively. Note that the 
upper energy bound on the absorption-like feature is not changed when the 
proton effects are neglected, but the shape of the feature differs 
considerably from the model which includes protons. In particular, lower 
energy bound of the feature moves to lower energies in the model without 
protons. The enhanced flux at lower energies in the model with protons 
occurs mainly because the absorption cross-section in the electron-proton 
plasma is significantly reduced over the pure electron plasma value below 
$\omega_{Bp}$ (see \S \ref{scat} and eq. \ref{abs}). 

\begin{figure}
\plotone{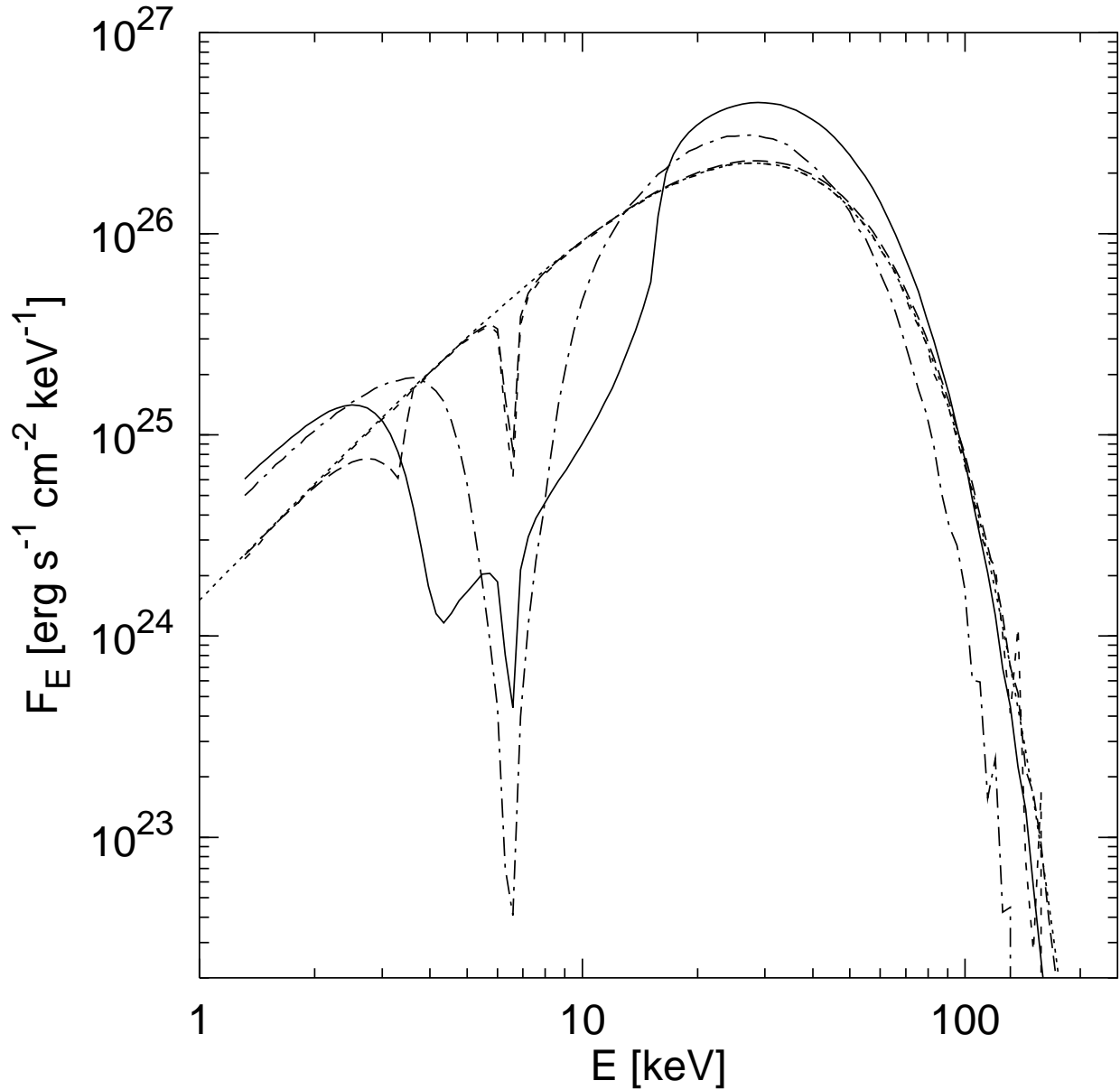}
\caption{Model spectra for the atmosphere with $H=10$\,cm and the 
magnetic field perpendicular to the surface in the case when vacuum 
polarization effects are turned off:  dot-dashed line is for $z=z_{0,2}$, 
and short-dashed line for $z=z_{0,1}$.  For reference, we also present  
the spectra from Fig. \ref{s10ep} for the model with proton and vacuum 
effects included.
\label{s10vac}}
\end{figure}

Figure \ref{s10vac} shows spectra for the model where vacuum polarization 
is neglected. For the production depth $z_{0,1}$ the flux is essentially 
blackbody apart from the proton line, which shape is not changed compared to 
the case with vacuum effects included. For photons injected at $z_{0,2}$,
the striking difference is visible in the proton absorption line, which is 
much broader and stronger when vacuum polarization is neglected. Note that 
in this case the proton line overlaps with the absorption-like feature in 
the spectrum which includes vacuum effects. The reduction in the proton 
line strength and width by vacuum polarization in such a case was also 
discussed  by \citet{2003MNRAS.338..233H} and \citet{2003ApJ...583..402O}. 

\begin{figure}
\plotone{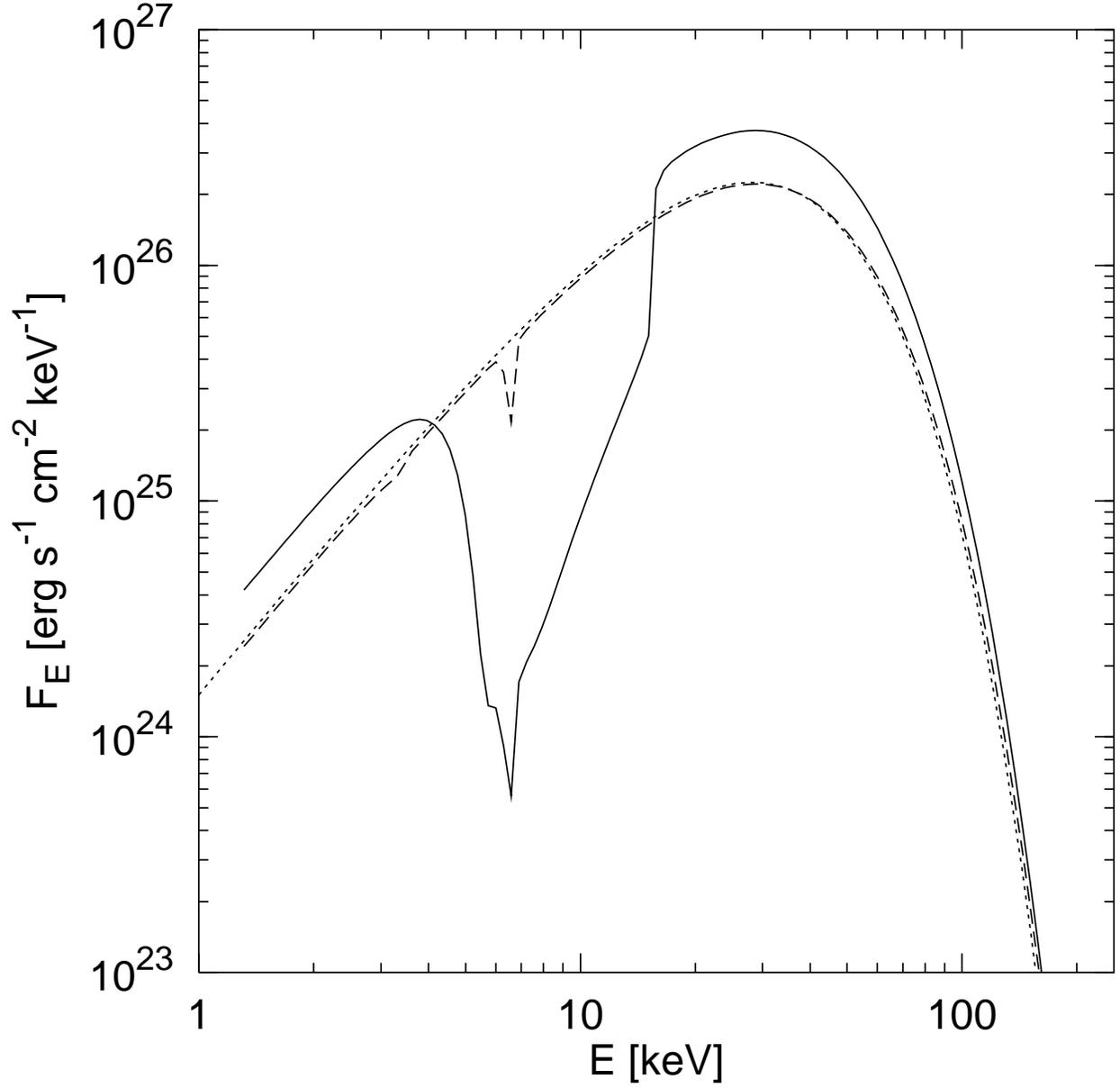}
\caption{Model spectra for the atmosphere with the 
magnetic field perpendicular to the surface and the smaller scaleheight 
$H=1$\,cm. In the case 
of the $z=z_{0,1}$ injection (long-dashed line), the vacuum feature 
nearly disappears, while for the $z=z_{0,2}$ case (solid line), the lower 
bound of the feature moves to higher energy compared with the model with
$H=10$\,cm.
\label{s1}}
\end{figure}

\begin{figure}
\plotone{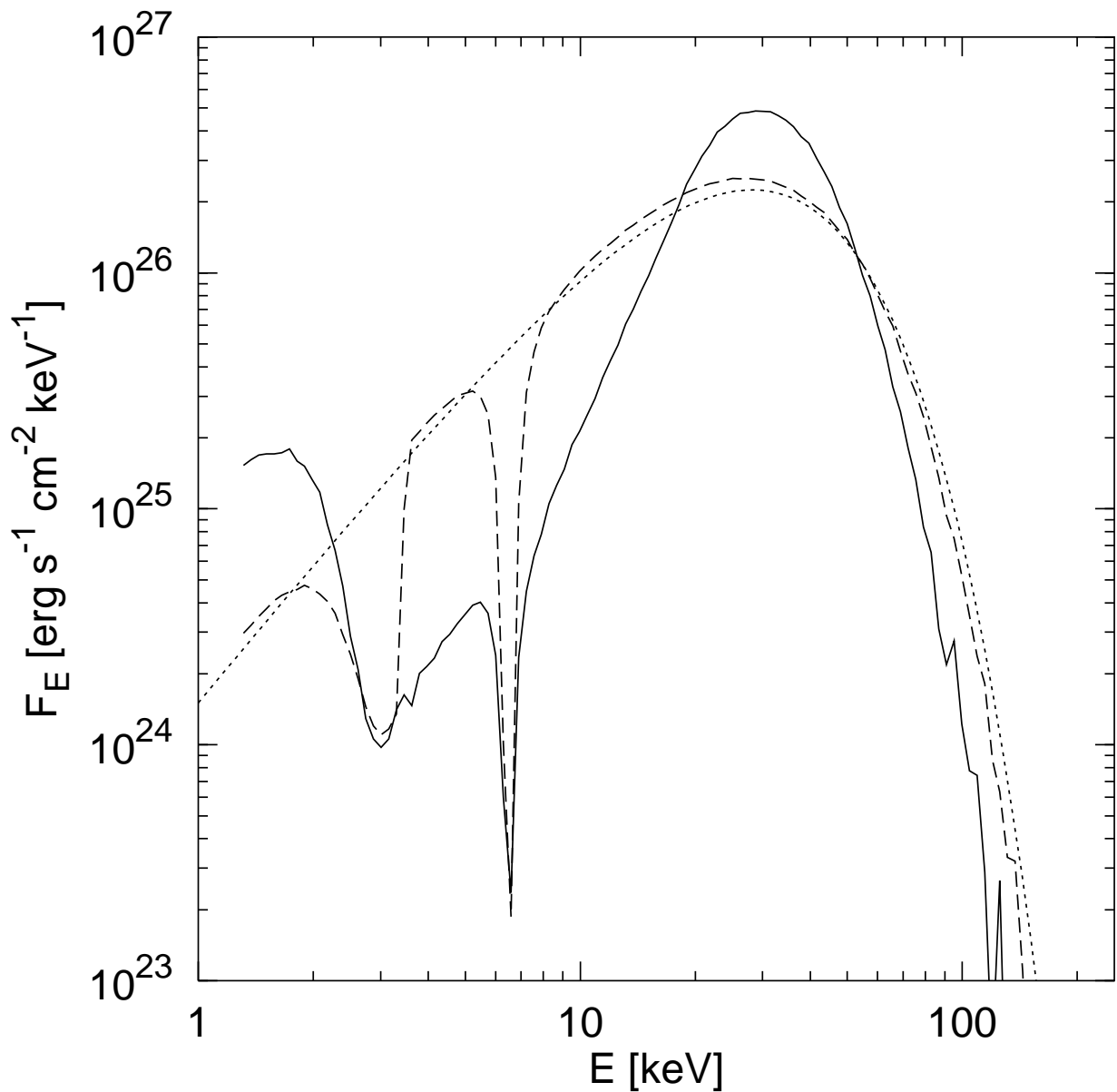}
\caption{Model spectra for the atmosphere with the 
magnetic field perpendicular to the surface and the larger scaleheight 
$H=100$\,cm. The slowly changing density gradients increase the influence of 
vacuum effects for both production depths: $z=z_{0,1}$ (long-dashed 
line) and  $z=z_{0,2}$ (solid line).
\label{s100}}
\end{figure}

As was noted in \citet{bul97}, the exact shape of the spectra and the 
vacuum feature depends also on other parameters than the injection depth,  
like the magnetic field strength, the scaleheight and the pair 
fraction. Figures \ref{s10ep} and \ref{s1}-\ref{s100} show the effects of 
changing the atmosphere scaleheight $H$. With  increasing $H$ the 
 vacuum feature broadens, the upper energy bound on the feature 
remains roughly constant, whereas the lower bound moves to lower energies. 
This is due to the fact that the typical density gradients are 
smoother when $H$ is large and hence the optical depth through
the vacuum resonance increases.
The depth of the proton cyclotron line also increases with scaleheight, 
which is especially marked for the lower production depth $z_{0,1}$. Note 
that in this case the vacuum feature and the proton line are 
considerably influenced by the value of $H$ and, in particular, 
they may be 
virtually nonexistent for low scaleheights (e.g. Fig. \ref{s1}). 

The spectra presented are essentially independent of viewing direction   
except for the angles near $\theta\approx 90^o$. In this case, the 
vacuum resonance disappears (see eqs \ref{q}-\ref{p}) which results 
in the lack of the absorption-like feature in the spectrum. Further, the 
depth of the proton line is considerably enhanced and the high-energy tail 
is reduced, which is due to geometric effects.

\begin{figure}
\plotone{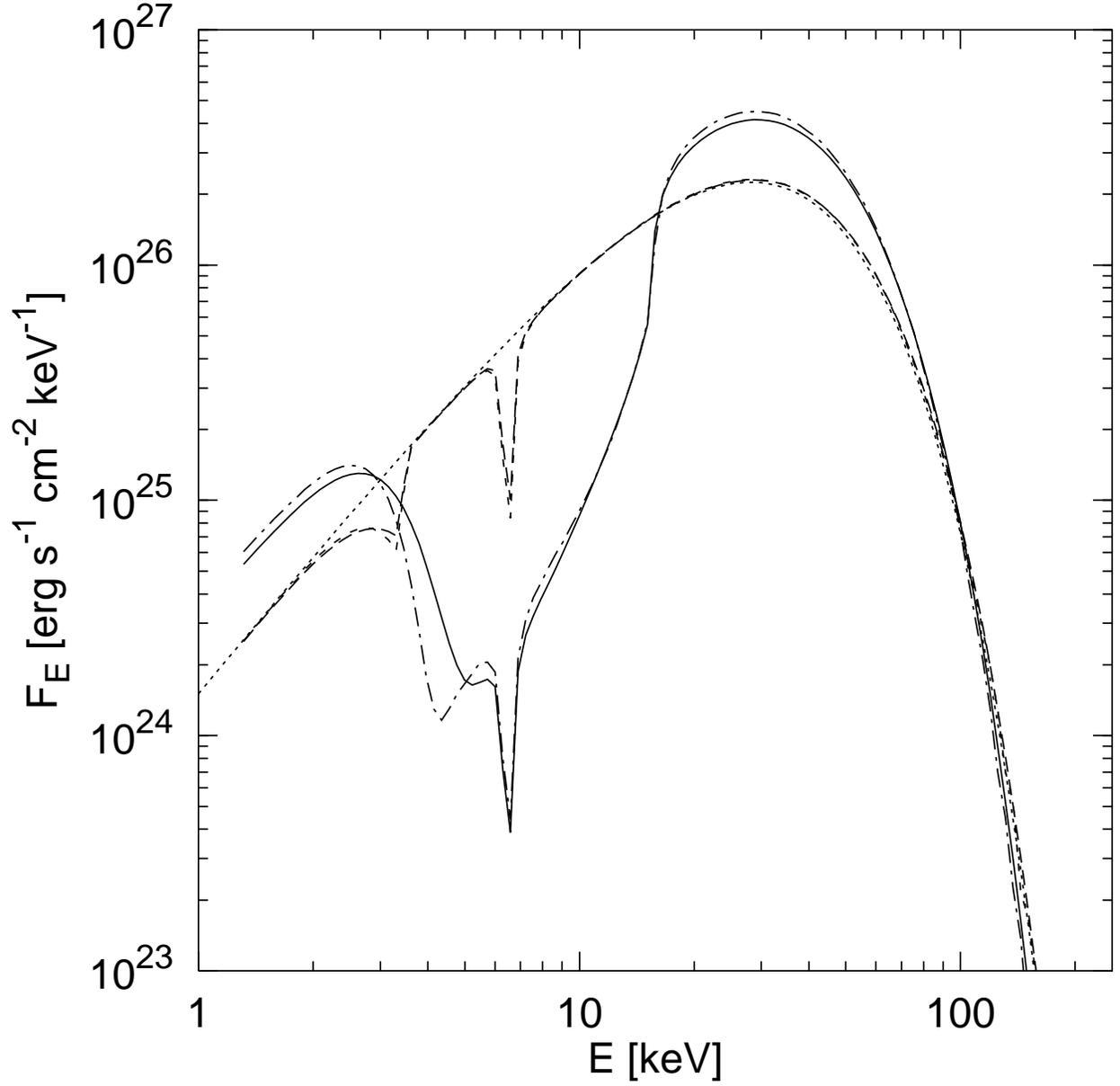}
\caption{Model spectra for the atmosphere with the 
magnetic field tilted at an angle of $45^\circ$ with respect to the 
surface normal and the scaleheight $H=10$\,cm. The $z=z_{0,1}$ case 
(long-dashed line) is practically indistinguishable from the model with 
perpendicular field (short-dashed line). For the larger injection 
depth, $z=z_{0,2}$ (solid line), the vacuum feature is slightly 
shifted to the higher energies, as compared to the case with perpendicular 
field (dot-dashed line).
\label{sobl}}
\end{figure}

Figure \ref{sobl} shows integrated photon spectra for the model with the 
magnetic field tilted at the angle $45^o$ to the neutron star surface 
($B=10^{15}$\,G) and the atmosphere scaleheight $H=10$\,cm. The overall 
shape of the spectrum does not significantly differ from the one  obtained 
in the model 
with perpendicular magnetic field, also shown in Figure \ref{sobl} for 
comparison. The angular dependence of radiation can be conveniently analyzed 
in a coordinate system $(\tilde{x},\tilde{y},\tilde{z})$ in which
the $\tilde{z}$-axis is directed along magnetic field.
In this system, the vacuum resonance disappears when 
$\tilde{\theta}=90^\circ$, where  $\tilde{\theta}$ is the wavevector 
inclination angle to the magnetic field. At a given propagation 
angle $\tilde{\theta}$ and growing azimuthal angle $\tilde{\phi}$
(the azimuthal angle is set to zero at the $-\tilde{y}$-axis and 
$\bf B$ lies in the $yz$-plane),
the depth 
of the proton line and the vacuum feature increases and the lower bound on 
the feature moves to lower energies. This is due to geometric effects, since 
at a given propagation angle $\tilde{\theta}$ and growing $\tilde{\phi}$, 
the physical depth to the top of the atmosphere increases.

\subsection{Polarization characteristics}
\label{pol}
Since the low-cross-section mode makes the main contribution to the 
photon flux emerging from the optically thick neutron star atmosphere, the 
observed radiation is expected to acquire strong linear polarization. As 
noted in \S \ref{npol}, the polarization pattern of this radiation is 
modified around the critical points in the system of vacuum and 
proton-electron plasma, which influence the frequency and angular 
dependences of normal mode polarization characteristics. 

\begin{figure}
\plotone{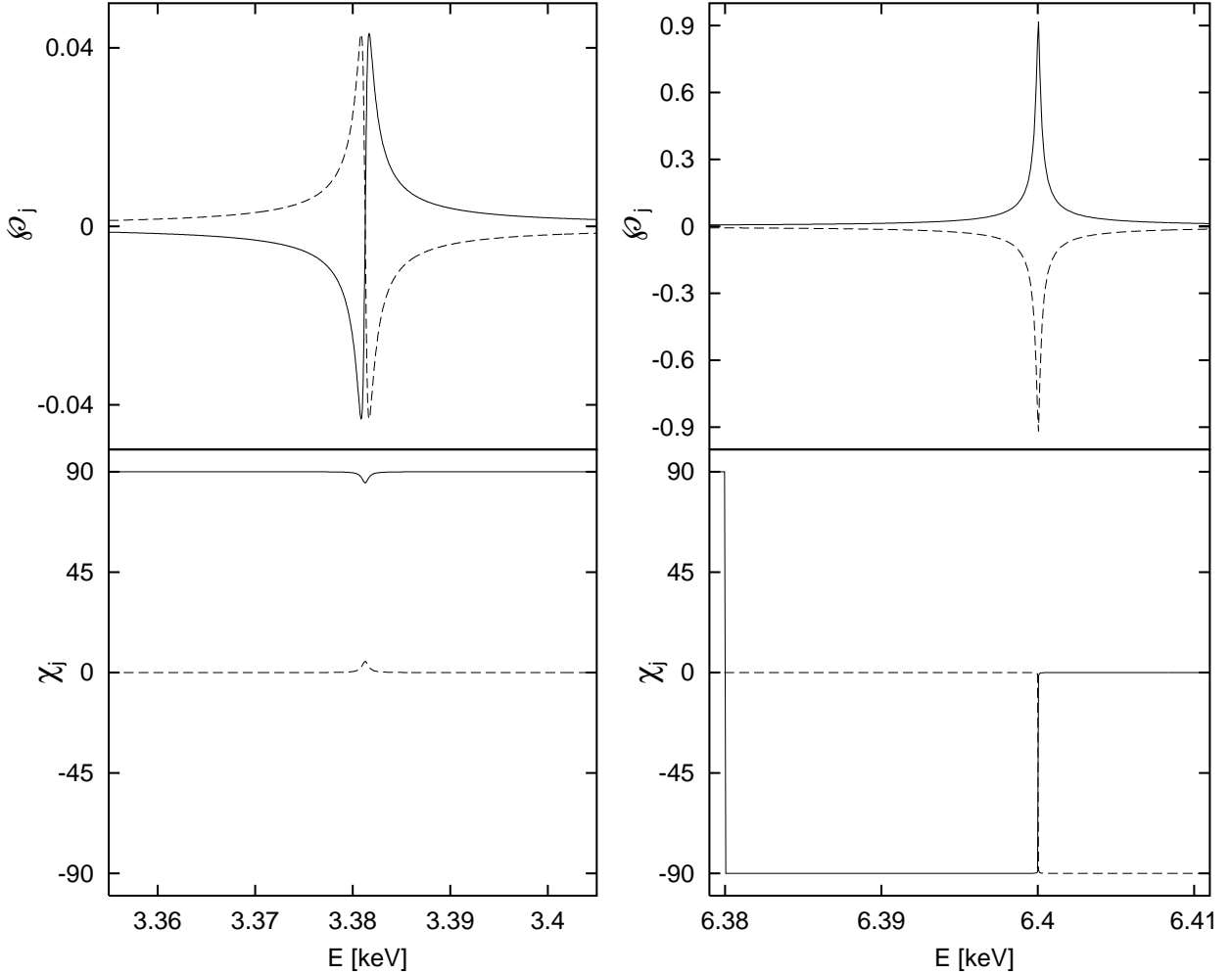}
\caption{The dependence of the normal mode ellipticities and  position 
angles on photon frequency near the vacuum resonance (left) and the
proton resonance (right). The extraordinary (low-cross-section) mode  
is the one with $\chi\approx\pi/2$. Thus, the labeling of the 
modes as extraordinary and ordinary does not follow the continuous lines.
The photon propagation angle is $\theta=78.5^\circ$ and $z=z_{0,1}$.
\label{elliptchi}}
\end{figure}

Figure \ref{elliptchi} shows an  example behavior of ellipticities ${\cal 
P}_j$ and position angles $\chi_j$ of the two normal modes in the vicinity 
 of the critical points in a plasma 
of density $\rho_{0,1}$ ($z=z_{0,1}$). 
The labeling of the modes is arbitrary.
The behavior of ${\cal P}_j$ and 
$\chi_j$ near the first critical frequency 
$\hbar\omega_{c,1}=\hbar\omega_{V,1}=3.38$ keV is different from the 
behavior at the second critical frequency 
$\hbar\omega_{c,2}=\hbar\omega_{Bp}=6.4$ keV, because for the  photon 
propagation direction we chose $\theta = 78.5^o$ ($\cos\theta=0.2$) $\theta 
> \theta_{c,1}=48^o$ ($\cos\theta_{c,1}=0.67$) but  $\theta < 
\theta_{c,2}=89.95^o$ ($\cos\theta_{c,2}=8\cdot 10^{-4}$) (see \S 
\ref{npol}). At the passage through the first critical frequency the sign of 
${\cal P}_j$ changes, but the position angles remain essentially the same. 
The modes are nearly linearly polarized (they become exactly linear at 
$\omega_{c,1}$) except around $\omega_{c,1}$, where they acquire weak 
elliptical polarization in the frequency range where $|q|$ is small and 
$|p|$ is close to unity. This also results in the small nonorthogonality of 
the modes around $\omega_{c,1}$. In the vicinity of the second critical 
frequency the normal mode polarization is strongly elliptical. The sign of 
${\cal P}_j$ remains the same but the position angles jump by $\pm\pi/2$ and 
coincide at $\omega_{c,2}$. Note also the jump by $\pi$ in the position 
angle of the low-cross-section mode at energy close to $\omega_{c,2}$. It 
occurs at the point where parameter $q$ diverges asymptotically ($|q|\gg 1$) 
and has opposite sign at both sides of the asymptote while at the same time 
$p$ crosses zero (compare eqs \ref{pj1}-\ref{chi1}). Note that the 
ellipticity is not influenced in this case (see also, e.g., \citet{pav79} 
and \citet{bul96} for a discussion of the energy dependence of the 
polarization characteristics). With growing density, $\theta_c$ 
corresponding to $\omega_{c}=\omega_{Bp}$ is approximately constant, whereas 
the critical angle corresponding to the vacuum resonance changes. However, 
at densities at which $\omega_{Bp}\approx\omega_V$ the critical angles 
disappear. 
  
\begin{figure}
\plotone{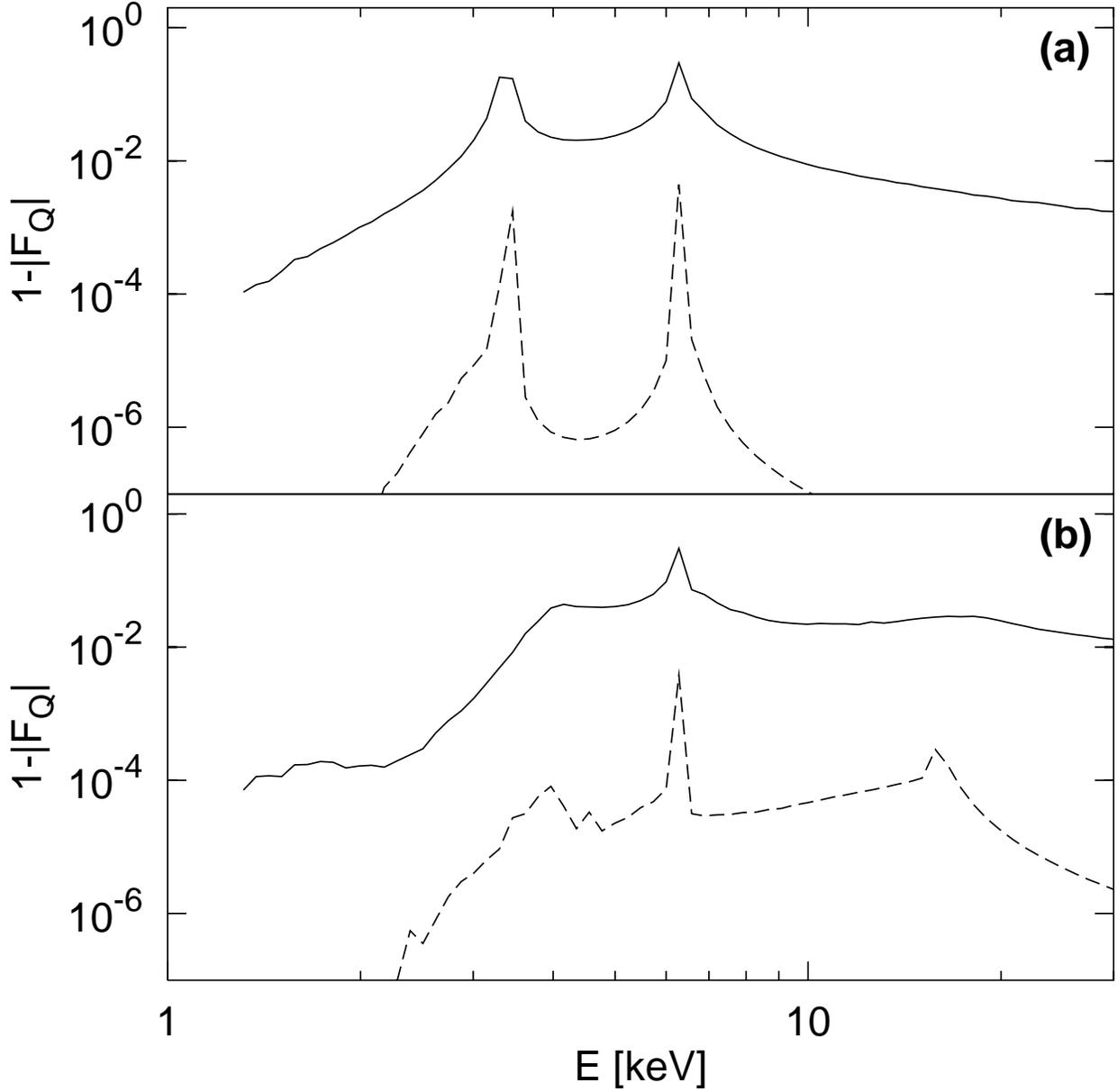}
\caption{The deviations from linear polarization,  $1-|F_Q|$, for the 
atmosphere with the magnetic field perpendicular to the surface and the 
standard scaleheight $H=10$\,cm for the injection depth  $z=z_{0,1}$ (a) 
and $z=z_{0,2}$ (b). The solid line corresponds to the propagation angle
$\theta\approx 8^\circ$, 
close to the direction of the magnetic field, 
and the dashed  line is for $\theta\approx 46^\circ$.
\label{polh10}}
\end{figure}

Figure \ref{polh10}  shows the variation of the linear polarization fraction
$F_Q=<Q>/I$  with photon frequency at the two  values of the 
propagation angle, for the atmosphere model with $H=10$\,cm, magnetic field 
perpendicular to the surface, and the production depths at $z=z_{0,1}$ 
(Fig. \ref{polh10}a)
and
$z=z_{0,2}$ (Fig. \ref{polh10}b), respectively. 
The radiation flux is essentially linearly polarized.
Small deviations from linearity 
  appear near the critical  frequencies  and especially for photons
propagating  nearly along the direction of the magnetic field.
For the case presented in Figure \ref{polh10}a, the 
decoupling depth is very close to the production depth $z_{0,1}$ (see Fig. 
\ref{zmean}) at all photon energies, and the location of the critical 
points is approximately as depicted in Figure \ref{elliptchi}.
Decoupling depths of photons in the model with the production depth 
located at $z=z_{0,2}$ span a wide range of $z$ between $\sim z_{0,1}$
and $z_{0,2}$, which spreads the deviation of polarization from linearity 
over the frequency range corresponding to the width of the 
vacuum feature.

\section{DISCUSSION}

We have presented a series of Monte Carlo 
calculations of the atmospheres in the $10^{15}$\,G 
magnetic field.  We took into account 
the electron scattering, the proton scattering, and the free-free 
absorption. The temperature of the atmosphere is taken to be constant
at $10$\,keV. The typical variations  of the temperature
with the optical depth in the difference scheme
calculations are weak, $T\approx \tau^{1/4}$ for $\tau\gg 1$
\citep{2001ApJ...563..276O,2001MNRAS.327.1081H}.
Therefore our assumption of uniform $T$ does not affect the results 
significantly. The main effect of including the temperature profile of this 
type would be to increase the flux at the high-energy part of the spectrum
\citep[see, e.g., for the emission in 
quiescence][]{2001MNRAS.327.1081H,2003MNRAS.338..233H}
We investigate in detail the effects of 
the vacuum resonance and the proton cyclotron resonance in the
spectra. We find that the critical frequencies in the
normal modes disappear at the density when the vacuum frequency 
overlaps the proton cyclotron resonance. 

In our treatment  of the radiative transfer we have neglected the 
effects of mode conversion  
\citep{2002ApJ...566..373L,2003ApJ...588..962L}. 
Mode conversion
may occur when  photon traverses a layer of varying density 
and encounters the vacuum resonance. At energies for which the 
vacuum resonance layer is optically thick in the extraordinary mode, the 
photon will be scattered or absorbed, which in our method is taken into 
account. Thus the mode conversion effects do not alter the results in this 
energy range. In the range of energies where the vacuum resonance
layer is  optically thin, the low-cross-section mode photon may  change its 
polarization to the high-cross-section one and be absorbed.
This effectively 
increases the optical depth through  the resonance layer. Thus the major 
effect of including the mode conversion effects would be to extend the 
vacuum absorption-like feature further to the low energies.
This effect has also been noted in \citet{2002ApJ...566..373L}.

We confirm the existence of a wide vacuum feature
in the spectra. \citet{bul97} suggested that the feature appears as a 
consequence of the Comptonization effects. We found that it is formed also 
in an absorption-dominated atmosphere, which means that the formation of the 
vacuum feature is a general property expected to occur in spectra of 
magnetar atmospheres.  
The vacuum feature appears if the plasma density 
at the photon production depth is large enough.
When the scaleheight is increased, the vacuum feature becomes more
prominent since the optical depth through 
the resonant layer increases (see the estimates in \citet{bul97}).
We have found that the approximate boundary of the density-scaleheight space 
region where the vacuum absorption-like feature appears in the spectrum 
can be estimated as $\rho_{inj}> 200 (H/1{\rm cm})^{-0.26}$g\,cm$^{-3}$,
where $\rho_{inj}$ is the density at which photons are injected into
the atmosphere.
Thus the vacuum feature
 should be prominent in magnetar spectra provided photons are produced
 at sufficiently large optical depths and the atmosphere scaleheight 
 is large enough. This second condition might be fulfilled for SGR 
bursts, since during the burst the atmosphere is probably lifted by 
radiation pressure and plasma is confined by the strong magnetic field.
We do not expect a significant absorption-like feature in the SGR spectra in 
quiescence. In this case the effective temperature is $T\approx 10^6$\,K and 
the vacuum feature would appear only as a broad depression in the 
exponentially falling high-energy tail, as shown in 
\citet{2003MNRAS.338..233H} and 
\citet{2001ApJ...563..276O,2003ApJ...583..402O}.

We find that the proton resonance appears as a narrow 
line on a wider background. This line is due to 
the proton resonance itself. The broad background line 
appears because the presence of the proton resonance affects the normal mode
polarization vectors in a broad frequency range around it,
which in turn influences the opacities.

We also calculate a model atmosphere with a magnetic field 
tilted at an angle of $45^\circ$ to the surface normal. The spectrum in
this case is very similar to the one obtained for the model with 
perpendicular magnetic field. Such field configurations are important when 
considering  radiation emitted from the entire surface of the star. Our 
result shows that the spectrum in the tilted field case can be well 
approximated by the the spectra calculated in the
simpler case of the field perpendicular to the surface. This is an 
independent confirmation of the earlier results obtained in  the diffusion 
approximation for the pulsar-type magnetic fields \citep[see, 
e.g.,][]{1995lns..conf...71P}. For the magnetar-type fields, the spectra for 
the field tilted at $90^\circ$ were shown to be similar to the $0^\circ$ 
case in the Monte Carlo simulations by \citet{bul97}, and in the diffusion 
approach by \citet{2001MNRAS.327.1081H}.

We present the polarization of the outgoing radiation.
It is essentially linear and the influence of the 
critical frequencies around the vacuum and proton 
resonances is very small. For the case of the small
production depth $z=z_{0,1}$, the effects of the vacuum resonance appear 
as deviations from the linear polarization near $\omega_V$, especially at 
the angles close to the direction of $\bf B$.  If the injection depth is 
large, $z=z_{0,2}$, the effects of the vacuum resonance are spread around
the width of the feature, as photons arrive to the 
observer from regions of different densities. 
The proton cyclotron resonance causes similar
distortions from linear polarization at $\omega_{Bp}$, regardless of the 
production depth.

SGR spectra in outbursts  have been first analyzed
in the hard X-ray range above $15$\,keV 
and the typical optically thin thermal bremsstrahlung (OTTB)
temperatures were
found to be in the range of $20$ to $40$\,keV \citep{2001ApJS..137..227A}.
The first analysis of the SGR spectra extending 
to $5$\,keV have been performed for the case of SGR 1806-20
 by \citet{1986Natur.322..152L} and \citet{1994ApJ...432..742F}.  
They found that the
 low energy spectrum was inconsistent with the extension
 of the OTTB spectra to the lower energies. 
 \citet{1998AIPC..428..947S} analyzed the RXTE observations
 of SGR 1806-20 and came up with a similar conclusion.
SGR 1900+14 was observed by HETE-2,
and \citet{2003AIPC..662...82O} have 
 shown that the low energy spectrum is also qualitatively similar.
It was not consistent  with an OTTB model, and was modeled
by the two blackbody spectra.
Recently \citet{2004ApJ...612..408F} have analyzed several 
bursts detected from SGR 1900+14 by Beppo SAX in 2001, and found that 
the OTTB model works fine above $15$\,keV but overestimates the flux
in the lower-energy range. They tried 
several alternative models and found the acceptable 
fits with the two blackbodies with temperatures
of $\sim 3$\,keV and $10$\,keV, and the respective radii
of $13$ and $2$\,km. We note that such a behavior of the spectrum resembles
the presence of a broad vacuum feature. Our model spectra
fall below the underlying thermal distribution around $\sim 10$\,keV. 
A double peaked spectrum as shown in, e.g., Figure~\ref{s100}
could well be modeled as two blackbodies with the temperatures
differing by a factor of a few. We suspect that these might be the hints
of the existence of the vacuum polarization feature 
in the spectra of SGR bursts. However, 
 in order to definitely claim the detection of
 such a feature one would have to perform the data analysis using the 
appropriate model spectra.

\acknowledgments
 We acknowledge support from KBN through grants PBZ-KBN-054/P03/2001 (JN, 
TB) and 2P03D 00125 (TB). TB is grateful for the hospitality of the Iowa 
State University, where this work was completed.

\newpage
\appendix

\section{PLASMA POLARIZATION TENSORS}
\label{app_1}
The general form of the plasma dielectric tensor $\epsilon_{ab}^{(pl)}$ 
which includes the contribution from both the scattering and free-free 
absorption processes was given by, e.g, \citet{pav79}. Rewriting their 
expressions in the coordinate system with the magnetic field along 
the $z$-axis, and the wavevector in the $yz$-plane at an angle $\theta$ to 
the magnetic field, we obtain for the plasma polarization tensors 
components ($s = e,p$ for electrons and protons):
\begin{eqnarray}
\label{app1}
\Pi_{xx}^{s} & = & 
\frac{\omega (\omega+{\rm i}\gamma_{\perp, s})}{(\omega+{\rm 
i}\gamma_{\perp, s})^2-\omega_{Bs}^2},\\
\Pi_{yy}^{s} & = & 
\frac{\omega (\omega+{\rm i}\gamma_{\perp, s})
[\omega+{\rm i}\gamma_{\parallel, s}\cos^2\theta 
(1+2\sin^2\theta) - {\rm i}\gamma_{\perp, s}\sin^2\theta\cos 2\theta]}
{(\omega+{\rm i}\gamma_{\parallel, s} )[(\omega+{\rm 
i}\gamma_{\perp, s})^2-\omega_{Bs}^2]},\\
\Pi_{zz}^{s} & = & 
\frac{\omega (\omega+{\rm i}\gamma_{\perp, s})
[\omega+{\rm i}\gamma_{\parallel, s}\cos^2\theta\cos 2\theta 
 + {\rm i}\gamma_{\perp, s}\sin^2\theta (1 + 2\cos^2\theta)]
-\omega\omega_{Bs}^2}{(\omega+{\rm i}\gamma_{\parallel, s} 
)[(\omega+{\rm i}\gamma_{\perp, s})^2-\omega_{Bs}^2]},\\
\Pi_{xy}^{s} & = & 
\frac{{\rm i} \omega\omega_{Bs}}{(\omega+{\rm 
i}\gamma_{\perp, s})^2-\omega_{Bs}^2},\\
\Pi_{xz}^{s} & = & 0,\\
\Pi_{yz}^{s} & = & 
\frac{-{\rm i} \omega (\omega+{\rm i}\gamma_{\perp, s})
(\gamma_{\perp,s} -\gamma_{\parallel, s})\sin 2\theta\cos2 \theta }
{2(\omega+{\rm i}\gamma_{\parallel, s} )[(\omega+{\rm i}\gamma_{\perp, 
s})^2-\omega_{Bs}^2]},
\end{eqnarray}
and 
\begin{equation}
\label{app2}
\Pi_{yx}^{s}=-\Pi_{xy}^{s},\;\;
\Pi_{zx}^{s}=\Pi_{xz}^{s},\;\; 
\Pi_{zy}^{s}=\Pi_{yz}^{s}.
\end{equation}
The above expressions take into account the influence the strong magnetic 
field has on the effective frequencies of electron-proton collisions, which 
then depend on photon polarization. Thus the longitudinal and transverse 
damping rates are
$\gamma_{\parallel, s} = \gamma_{\parallel, cs} + \gamma_{rs}$ and
$\gamma_{\perp, s} = \gamma_{\perp, cs} + \gamma_{rs}$, respectively, 
where the radiative widths $\gamma_{rs}= (2/3)(e^2/m_s c^3)\omega^2$ and 
the corresponding collision frequencies 
$\gamma_{\parallel, cs} = (\sigma_a g_{\parallel}/\sigma_T)\gamma_{rs}$ 
and  $\gamma_{\perp, cs} = (\sigma_a g_{\perp}/\sigma_T)\gamma_{rs}$
\citep[see also][]{pav76},
with $\sigma_a$ given by equation (\ref{siga}) and
magnetic Gaunt factors $g_{\parallel, \perp}$. 
When only scattering processes are considered,  
$\gamma_{\parallel, s}=\gamma_{\perp, s}=\gamma_{rs}$, plasma 
polarization tensors become diagonal in the rotating coordinates and take 
the form as given by equation (\ref{pialfa}).



\clearpage

\end{document}